\documentclass[prd,nofootinbib,english,a4paper]{revtex4-2}
\usepackage{amsmath}
\usepackage{amsfonts,color}
\usepackage{amsmath}
\usepackage{amsthm}
\usepackage{amsfonts,color,xcolor}
\usepackage{amssymb,float}
\usepackage{amsmath}
\usepackage{amsfonts}
\usepackage{amssymb}
\usepackage{amsthm}
\usepackage{accents}
\usepackage{mathrsfs}
\usepackage{accents}
\usepackage{hyperref}
\hypersetup{
	colorlinks=true,
	linkcolor=blue,
	filecolor=magenta,
	citecolor=blue
}
\usepackage[utf8]{inputenc}
\usepackage{graphicx}
\allowdisplaybreaks

\newcommand{\udt}[3]{#1^{#2}_{\phantom{#2}#3}}

\newcommand{\dut}[3]{#1_{#2}^{\phantom{#2}#3}}

\DeclareMathOperator{\curl}{curl}

\usepackage{bookmark}

\newcommand{\dd}{\mathrm{d}}

\newcommand{\lc}[1]{\accentset{\circ}{#1}}
\newcommand{\torvec}{\mathscr{V}}
\newcommand{\toraxi}{\mathscr{A}}
\newcommand{\ginv}[2]{\underaccent{\mathsf{#2}}{\hat{\boldsymbol{#1}}}}
\newcommand{\gdep}[2]{\underaccent{\mathsf{#2}}{\hat{#1}}}
\newcommand{\vek}[1]{\underline{#1}}
\newcommand{\mat}[1]{\underline{\underline{#1}}}

\begin{document}

\title{Perturbations in Non-Flat Cosmology for \texorpdfstring{$f(T)$}{f(T)} gravity}

\author{Sebastian Bahamonde}
\email{sbahamondebeltran@gmail.com, bahamonde.s.aa@m.titech.ac.jp}
\affiliation{Department of Physics, Tokyo Institute of Technology
1-12-1 Ookayama, Meguro-ku, Tokyo 152-8551, Japan.}
\affiliation{Laboratory of Theoretical Physics, Institute of Physics, University of Tartu, W. Ostwaldi 1, 50411 Tartu, Estonia}

\author{Konstantinos F.	Dialektopoulos}
\email{kdialekt@gmail.com}
\affiliation{Laboratory of Physics, Faculty of Engineering, Aristotle University of Thessaloniki, Thessaloniki 54124, Greece}
\affiliation{Department of Physics, Nazarbayev University, 53 Kabanbay Batyr avenue, 010000 Astana, Kazakhstan}

\author{Manuel Hohmann}
\email{manuel.hohmann@ut.ee}
\affiliation{Laboratory of Theoretical Physics, Institute of Physics, University of Tartu, W. Ostwaldi 1, 50411 Tartu, Estonia}

\author{Jackson Levi Said}
\email{jackson.said@um.edu.mt}
\affiliation{Institute of Space Sciences and Astronomy, University of Malta, Msida, Malta}
\affiliation{Department of Physics, University of Malta, Msida, Malta}

\author{Christian Pfeifer}
\email{christian.pfeifer@zarm.uni-bremen.de}
\affiliation{ZARM, University of Bremen, 28359 Bremen, Germany}

\author{Emmanuel N. Saridakis}
\email{msaridak@noa.gr}
\affiliation{National Observatory of Athens, Lofos Nymfon, 11852 Athens, Greece}
\affiliation{CAS Key Laboratory for Research in Galaxies and Cosmology,University of Science and Technology of China, Hefei, Anhui 230026, China}

\begin{abstract}
The study of cosmological perturbation theory in $f(T)$ gravity is a topic of great interest in teleparallel gravity since this is one of the simplest generalizations of the theory that modifies the teleparallel equivalent of general relativity. In this work, we explore the possibility of a non-flat FLRW background solution and perform perturbations for positively as well as negatively curved spatial geometries, together with a comparison to the flat case. We determine the generalized behaviour of the perturbative modes for this non-flat FLRW setting for arbitrary $f(T)$ models, when the most general homogeneous and isotropic background tetrads are used. We also identify propagating modes in this setup, and relate this with the case of a flat cosmology.
\end{abstract}

\maketitle

\section{Introduction}

Over the last several decades, the Universe has not only been measured to be accelerating \cite{SupernovaSearchTeam:1998fmf,SupernovaCosmologyProject:1998vns} but to be expanding faster than what would be expected using the $\Lambda$CDM concordance model \cite{DiValentino:2021izs}. The most striking disagreement is highlighted in the so-called Hubble tension. Here, model independent measures of the Hubble constant $H_0$ from local measurements provide higher Hubble constant values, such as from the SH0ES \cite{Riess:2021jrx} and H0LiCOW collaborations \cite{Wong:2019kwg}, when compared with flat $\Lambda$CDM predictions from the early Universe, such as from the Planck Collaboration \cite{Planck:2018vyg} or Dark Energy Survey \cite{DES:2021wwk}. The broader spectrum of cosmological tensions \cite{DiValentino:2020vhf,DiValentino:2020zio,DiValentino:2020vvd} has prompted a revival in theories beyond general relativity (GR) with a renewed interest in the literature in precision tests of these theories.

Teleparallel geometry \cite{Aldrovandi:2013wha} offers an alternative framework to construct gravitational theories in addition to curvature-based geometries \cite{misner1973gravitation}. This is achieved by using a teleparallel connection $\Gamma^{\sigma}{}_{\mu\nu}$ \cite{Bahamonde:2021gfp,Krssak:2018ywd,Cai:2015emx}, which possesses torsion and has vanishing curvature, instead of the Levi-Civita connection $\lc{\Gamma}^{\sigma}{}_{\mu\nu}$ of a metric, which is torsion free but yields a non-vanishing curvature of spacetime; both connections are metric compatible (here and in what follows, over-circles are used to denote quantities that are calculated using the Levi-Civita connection). The result is an altogether novel composition of gravitational theories. Also, teleparallel gravity was first used to construct a teleparallel equivalent of general relativity (TEGR) \cite{Maluf:2013gaa,aldrovandi1995introduction} which is dynamically equivalent to GR. This means that TEGR and GR agree on all classical tests but may differ when considering non-classical regimes. Thus, the same evidence for GR also supports its teleparallel formulation TEGR.

As in curvature-based theories of gravity \cite{CANTATA:2021ktz,Addazi:2021xuf}, teleparallel gravity (TG) theories have emerged in various forms beyond TEGR \cite{Bahamonde:2021gfp,Krssak:2018ywd,Cai:2015emx}. The most natural extension to the TEGR action, defined by the Torsion scalar $T$, is $f(T)$ gravity where the Lagrangian is an arbitrary function of the TEGR Lagrangian. Analogous to $f(\lc{R})$ gravity, $f(T)$ gravity provides different avenue to confronting the observational and theoretical challenges of $\Lambda$CDM cosmology. Unlike $f(\lc{R})$ gravity, $f(T)$ gravity is generically second-order in derivatives and so offers an intriguing platform to constructing gravitational models that are not exposed to possible unhealthy features resulting from high-order derivative theories. Teleparallel geometry is also the basis for other manifestations of teleparallel gravity theories such as New General Relativity \cite{Hayashi:1979qx,Mikhail:1996hw}, $f(T,B)$ gravity \cite{Bahamonde:2015zma} ($B$ represents the difference between the Ricci and torsion scalars and is a boundary term) and $f(T,T_G)$ gravity \cite{Kofinas:2014aka,Kofinas:2014daa,Kofinas:2014owa} ($T_G$ represents the teleparallel analogue of the Gauss-Bonnet term). There has also been a significant amount of work exploring possible scalar-tensor extensions of TG \cite{Bahamonde:2022lvh,Bahamonde:2019shr,Bahamonde:2019ipm,Bahamonde:2020cfv,Bahamonde:2021dqn,Dialektopoulos:2021ryi,Hohmann:2018vle,Hohmann:2018dqh,Hohmann:2018ijr} including the coupling of pseudo-scalars (axions) \cite{Li:2020xjt,Hohmann:2020dgy}.

Teleparallel and curvature-based geometries differ in more than just by the choice of the connection. TG tends to produce theories which have an explicit appearance of the local Lorentz frame in the ensuing field equations \cite{Aldrovandi:2013wha,Hehl:1994ue}. To maintain Lorentz invariance requires the addition of an dynamical spin connection $\udt{\omega}{A}{B\mu}$ in the gravitational action. The corresponding six additional field equations determine six degrees of freedom associated to Lorentz transformations \cite{Krssak:2015oua}.

The aforementioned local Lorentz invariance implies that locally it is always possible to find a Lorentz transformation such that after this transformation the spin connection coefficients vanish; this Lorentz gauge choice is called the Weitzenb\"{o}ck gauge. Choosing this gauge, the six Lorentz degrees of freedom are contained purely in the tetrad field, which is then determined by the field equations. Neglecting this fact would severely constrain the applicability of the teleparallel geometry~\cite{Tamanini:2012hg,Krssak:2018ywd}. A constructive approach how to obtain the Weitzenb\"{o}ck gauge follows from its geometric interpretation~\cite{Hohmann:2021dhr}.

The covariant formulation of theories in TG can thus provide a suitable base on which to study gravitational systems. In the TEGR formulation of TG, all solutions appear in the Weitzenb\"{o}ck gauge due to the form of the theory. Moreover, they also feature healthy perturbations at all orders \cite{Bahamonde:2021gfp}. However, perturbations in $f(T)$ gravity about some solutions have led to higher perturbative order terms interacting with linear perturbations thus showing strong coupling for such spacetimes \cite{Golovnev:2020zpv}. This has led to serious concerns about the perturbative structure of $f(T)$ cosmology about the flat Friedmann-Lema\^{i}tre-Robertson-Walker (FLRW) cosmology \cite{Golovnev:2018wbh}. The issue has also been found in perturbations about Minkowski spacetimes \cite{Golovnev:2020nln,Jimenez:2020ofm}. The source of the problem is related with the degrees of freedom of the theory not appearing at linear order in these solutions \cite{Blagojevic:2020dyq,Guzman:2019oth}. The issue has not been studied beyond $f(T)$ gravity but may appear in some branches of these theories.

It is hence crucial that the scale and scope of possible strongly coupled solutions be further studied in $f(T)$ gravity to better understand their impact on physical settings. In this work, we explore cosmological perturbations about a non-flat FLRW cosmological background \cite{Paliathanasis:2021uvd} to assess whether it is strongly coupled. This is important not only to examine whether strong coupling has also infiltrated non-flat cosmologies but also due to recently renewed interested in these cosmologies \cite{DiValentino:2019qzk,Handley:2019tkm}. We do this by first briefly reviewing the foundations of $f(T)$ gravity in Sec.~\ref{sec:f_T_Intro} which is then expanded to include the perturbation strategy in Sec.~\ref{sec:f_T_perturb}. The main results are contained in Sec.~\ref{sec:cosmo_perturb} where we present the cosmological perturbations about a non-flat FLRW background. In Sec.~\ref{sec:conclu} we conclude with a discussion of our main results and the issue of strong coupling in this setting.

\section{Teleparallel gravity and \texorpdfstring{$f(T)$}{f(T)}}\label{sec:f_T_Intro}

Teleparallel gravity is solely based on the torsion of the connection of spacetime, in contrast with GR which is based on its curvature. This is done by replacing the Levi-Civita connection with the Teleparallel one and thus, we end up with a new framework for the gravitational interactions with which we can construct new theories of gravity. The gravitational field in curvature based theories is measured by the Riemann tensor and its contractions; in Teleparallel theories the curvature and thus the Riemann tensor itself, vanish identically and gravity is measured through torsion\footnote{Note that, the Riemann tensor calculated with the teleparallel connection vanishes, while the one calculated with the Levi-Civita connection does not.}.

The dynamical variable in General Relativity is the metric tensor, $g_{\mu\nu}$, and it encodes all the necessary information for the gravitational field, since using it we can calculate not only the Levi-Civita connection but also the Riemann tensor. In the Teleparallel framework the metric is substituted by a tetrad-spin connection pair, $\{e^A{}_{\mu},\omega ^A{}_{B\mu}\}$, where the Greek indices denote coordinates on the general manifold and Latin indices on the local Minkowski space, $\eta _{AB}$, where $\eta_{AB}=\mathrm{diag}(-,+,+,+)$. The tetrad $e^A{}_{\mu}$ (inverse tetrad $E_A{}^{\mu}$) is used to raise Minkowski space (general manifold) indices to the general manifold (Minkowski space) through the relations
\begin{equation}\label{eq:metric}
    g_{\mu\nu} = e^A{}_\mu e^B{}_{\nu} \eta _{AB} \quad \text{and} \quad \eta_{AB} = e_A{}^\mu e_B{}^\nu g_{\mu\nu}\,,
\end{equation}
and they also satisfy the orthogonality conditions
\begin{equation}
    e^A{}_\mu e_B{}^\mu = \delta ^A_B \quad \text{and} \quad e^A{}_{\mu}e_A{}^\nu = \delta _\mu ^\nu \,.
\end{equation}
The local Lorentz transformations (LLTs) on the local Minkowski space, $\Lambda ^A{}_B$, dictate that the tetrad has 6 extra degrees of freedom (DoFs) compared to the metric and thus different tetrad can reproduce the same metric.

GR can be expressed also in terms of the tetrad formulation \cite{Hohmann:2021fpr}, however it is not so common. The teleparallel connection though, that has no curvature and is metric compatible, is expressed in terms of the tetrad and the spin connection as
\begin{equation}
\Gamma^{\lambda}{}_{\nu\mu}=\dut{e}{A}{\lambda}\partial_{\mu}\udt{e}{A}{\nu}+\dut{e}{A}{\lambda}\udt{\omega}{A}{B\mu}\udt{e}{B}{\nu}\,,
\end{equation}
where $\omega ^A{}_{B\mu}$ is a flat spin connection that satisfies
\begin{equation}
    \partial_{[\mu}\udt{\omega}{A}{|B|\nu]} + \udt{\omega}{A}{C[\mu}\udt{\omega}{C}{|B|\nu]} \equiv 0\,.
\end{equation}
The spin connection is flat and metric compatible and the theory remains covariant. It is always possible to choose a gauge such that this spin connection vanishes identically, $\omega ^A{}_{B\mu} = 0$; this gauge is called Weitzenb\"ock gauge. In any other gauge the spin connection can be written as
\begin{equation}
    \omega ^A{}_{B\mu} = \Lambda ^A{}_C \partial _\mu (\Lambda ^{-1})^C{}_B \,,
\end{equation}
and thus it is a pure gauge DoF.

As already discussed, gravity in the teleparallel framework is mediated through torsion. Hence, it would be helpful if we defined a tensor to replace the curvature tensor used in GR. This torsion tensor is the antisymmetric part of the connection
\begin{equation}\label{eq:tor_ten}
    T^A{}_{\mu\nu} = 2 \Gamma ^A{}_{[\nu\mu]}\,.
\end{equation}
The scalar that appears in the action of the Telleparallel Equivalent of General Relativity is called the torsion scalar and is defined as
\begin{equation}
    T= \frac{1}{4}T^{\rho\sigma\mu}T_{\rho\sigma \mu} + \frac{1}{2}T^{\mu\sigma\rho}T_{\rho\sigma\mu} - T^\rho{}_{\rho\sigma}T^\mu{}_\mu{}^{\sigma}.
\end{equation}
Its relation with the Ricci scalar, calculated with the Levi-Civita connection is given by
\begin{equation}\label{ricci_scalar}
    \lc{R} = - T+B\,,
\end{equation}
where $B$ is a total divergence term define as
\begin{equation}
    B = \frac{2}{e} \partial _\rho (eT^\mu {}_\mu{}^\rho)\,,
\end{equation}
where $e$ is the determinant of the tetrad $e^A{}_\mu$. It should be noted once again that the total curvature of the spacetime, meaning the Ricci scalar calculated with the general teleparallel connection vanishes identically, i.e. $R =\lc{R} + T - B = 0$. Overcircles refer to quantities computed with the Levi-Civita connection.

From Eq.~\eqref{ricci_scalar} it can be realized that at the level of the action, the only surviving term in the functional integral will be the torsion scalar, since the total divergence term will not contribute. Thus, at the level of equations, the two theories are equivalent. An interesting characteristic of the TEGR is that it can be formulated as a gauge theory of the translation group. The action of the theory in the presence of matter reads
\begin{equation}
S = S_{\rm g} + S_{\rm m}= -\frac{1}{2\kappa^2}\int T\,e\,\dd^4x + \int \mathcal{L}_{\rm m}e \,\dd^4 x\,,
\end{equation}
and the associated equations of motion are derived by varying the action with respect to the tetrad $e^A{}_\mu$ and are
\begin{equation}
E_A{}^{\mu}=\frac{1}{\kappa^2}\left[\frac{1}{e}\partial _{\sigma}(eS_A{}^{\mu\sigma})-T^\sigma{}_{\nu A}S_\sigma {}^{\nu\mu} + \frac{1}{2}e_A{}^\mu T + \omega ^B{}_{A \nu}S_B{}^{\nu\mu}\right] = \Theta _A{}^\mu\,,
\end{equation}
where $E_A{}^\mu$ is the variation of the pure gravitational Lagrangian with respect to the tetrad (including the gravitational constant $\kappa$) and $\Theta _A{}^{\mu}$ is the energy-momentum tensor defined as $\Theta _A{}^{\mu} = \frac{1}{e}\frac{\delta (e \mathcal{L}_{\text{m}})}{\delta e ^A{}_{\mu}}$\,.

As discussed in the introduction, a plethora of modifications beyond TEGR has been proposed in the literature, with the most straightforward extension being the so-called $f(T)$ gravity, that is a generalization of the torsion scalar in the action, to an arbitrary function of it. Its action reads,
\begin{equation}\label{fT_action}
S = -\frac{1}{2\kappa^2}\int f(T)\,e\,\dd^4x + S_{\text{m}}\,
\end{equation}
and by varying the action with respect to the tetrad $e^A{}_\mu$ we get its equations of motion, which expressed in general manifold's indices are
\begin{equation}\label{eq:ftEOM}
E_{\mu\nu} =\frac{1}{\kappa^2}\left[-\frac{1}{2}fg_{\mu\nu} + S^{\rho\sigma}{}_{\mu}(T_{\rho\sigma\nu} - K_{\rho\nu\sigma})f_T - \lc{\nabla}_{\rho}(S_{\nu\mu}{}^{\rho}f_T)\right] = \Theta_{\mu\nu}\,,
\end{equation}
with $f_T = \dd f(T)/\dd T$. In this representation of the field equations the symmetric part of the energy momentum tensor $\Theta_{(\mu\nu)}$ is the Hilbert energy-momentum tensor, which is the source of the gravitational dynamics in theories of gravity based on a spacetime metric. Modified teleparallel theories have both the tetrad and the spin connection as their field variables. However, variations of the action \eqref{fT_action} with respect to the spin connection $\omega ^A{}_{B\mu}$ would just lead to the antisymmetric part of the field equations of the tetrad, which have a vanishing source, $\Theta_{[\mu\nu]}=0$, for the matter coupling we assume here. That is, we consider a matter coupling for which matter only couples to the tetrad (metric) and not to the connection. One could also choose to couple the matter to the teleparallel spin connection, a thorough discussion of these options can for example be found in~\cite{BeltranJimenez:2020sih}.

\section{Cosmological perturbations: basic ingredients}\label{sec:f_T_perturb}

In this section, we recall the framework of cosmological perturbation theory in teleparallel gravity, before we apply it to $f(T)$-gravity in Sec.~\ref{sec:cosmo_perturb}. We introduce the most general spatially homogeneous and isotropic tetrads and tensors needed. Moreover we discuss the formalism how to study the cosmological perturbations in teleparallel gravity most conveniently, i.e. a $3+1$ decomposition of the dynamical fields adapted to the symmetry of the homogeneous and isotropic background solution, the form of the resulting perturbed field equations for any teleparallel theory of gravity, as well as the emergence of coordinate gauge transformations and suitable harmonic expansions of the degrees of freedom.

\subsection{Cosmologically symmetric tetrads and tensors}
A spacetime manifold possesses a certain symmetry, if the tensors which define the geometry of the manifold are invariant under a group of diffeomorphisms. For teleparallel gravity the notion of symmetry as been discussed for example in Refs.~\cite{Hohmann:2019nat, Pfeifer:2022txm}.

The most general spatially homogeneous and isotropic teleparallel geometry in the  Weitzenb\"ock gauge is given by two branches of tetrads \cite{Hohmann:2020zre}, the so called vector and axial branch. We denote the curvature parameter of the spatial homogeneous and isotropic spaces by $u=\sqrt{k}$.

The vector branch tetrad is given by
\begin{equation}\label{eq:vectet}
e^A{}_\mu=\left(
\begin{array}{cccc}
N(t)\sqrt{1-r^2 u^2} & \frac{i r u a(t)}{\sqrt{1-r^2 u^2}} & 0 & 0 \\
i r u N(t) \sin\vartheta \cos\varphi & a(t) \sin\vartheta \cos\varphi & r a(t) \cos\vartheta \cos\varphi & -r a(t) \sin\vartheta \sin\varphi \\
i r u N(t) \sin\vartheta \sin\varphi & a(t) \sin\vartheta \sin\varphi & r a(t) \cos\vartheta \sin\varphi & r a(t) \sin\vartheta \cos\varphi \\
i r u N(t) \cos\vartheta & a(t) \cos\vartheta & -r a(t) \sin\vartheta & 0 \\
\end{array}
\right)\,,
\end{equation}
while the so called axial branch tetrad is
\begin{equation}\label{eq:axtet}
e^A{}_\mu=\left(
\begin{array}{cccc}
N(t) & 0 & 0 & 0 \\
0 & \frac{a(t) \sin\vartheta \cos\varphi}{\sqrt{1-r^2 u^2}} & r a(t) \left(\cos\vartheta \sqrt{1-r^2 u^2} \cos\varphi+r u \sin\varphi\right) & r a(t) \sin\vartheta \left(r u \cos\vartheta \cos\varphi-\sqrt{1-r^2 u^2} \sin\varphi\right) \\
0 & \frac{a(t) \sin\vartheta \sin\varphi}{\sqrt{1-r^2 u^2}} & r a(t) \left(\cos\vartheta \sqrt{1-r^2 u^2} \sin\varphi-r u \cos\varphi\right) & r a(t) \sin\vartheta \left(\sqrt{1-r^2 u^2} \cos\varphi+r u \cos\vartheta \sin\varphi\right) \\
0 & \frac{a(t) \cos\vartheta}{\sqrt{1-r^2 u^2}} & -r a(t) \sin\vartheta \sqrt{1-r^2 u^2} & -r^2 u a(t) \sin ^2\vartheta \\
\end{array}
\right)\,.
\end{equation}
Both tetrads yield via \eqref{eq:metric} the standard homogeneous and isotropic metric
\begin{equation}\label{eq:cosmomet}
    \dd s^2 = g_{\mu\nu}\dd x^\mu \dd x^\nu =-N(t)^2\dd t^2+a(t)^2\Big[\frac{\dd r^2}{1-u^2r^2}+r^2\dd\Omega^2\Big]\,.
\end{equation}
The torsion $T^\rho{}_{\mu\nu}$ generated by these tetrads can be displayed most conveniently by introducing a 3+1-decomposition of the metric \cite{Hohmann:2020vcv}, as
\begin{align}
    g_{\mu\nu} = -n_{\mu}n_{\mu} + h_{\mu\nu}\,,
\end{align}
where the conormal $n_\nu$ to the spatial hypersurfaces and the spatial metric $h_{\nu\mu}$, suppressing the explicit time dependence of the function $N$ and $a$, are given by
\begin{align}\label{eq:nh}
    n_\mu = (-N,0,0,0)\,,\quad h_{\mu\nu} = \mathrm{diag}\left(0,\frac{a^2}{1-u^2 r^2},a^2 r^2, a^2 r^2 \sin^2\vartheta\right)\,.
\end{align}
Moreover we need the totally antisymmetric Levi-Civita tensors $\epsilon_{\mu\nu\rho\sigma}$ of the spacetime metric $g$ and $\varepsilon_{\mu\nu\rho} = n^\sigma \epsilon_{\sigma\mu\nu\rho}$ of the spatial metric $h$, which are defined through
\begin{align}
    \epsilon_{tr\vartheta\varphi} = \frac{N a^3r^2\sin\vartheta}{\sqrt{1-u^2 r^2}} \,,\quad \epsilon_{r\vartheta\varphi}=\frac{a^3r^2\sin\vartheta}{\sqrt{1-u^2 r^2}}\,.
\end{align}
From now on, we will choose the conformal time gauge with $N(t)=a(t)$. The torsion~\eqref{eq:tor_ten} of the spatially homogeneous and isotropic tetrads is defined in terms of two functions $\torvec$ and $\toraxi$, namely \cite{Iosifidis:2020gth}
\begin{equation}\label{eq:cosmotor}
T^{\mu}{}_{\nu\rho} = \frac{2}{a
}\Big(\torvec h^{\mu}_{[\nu}n_{\rho]} + \toraxi\varepsilon^{\mu}{}_{\nu\rho}\Big)\,.
\end{equation}
For the vector branch tetrad \eqref{eq:vectet} we find
\begin{equation}\label{eq:vectorbranch}
\torvec = \mathcal{H} \pm iu\,, \quad \toraxi = 0\,,
\end{equation}
while for the axial ranch tetrad \eqref{eq:axtet} we obtain
\begin{equation}\label{eq:axialbranch}
\torvec = \mathcal{H}\,, \quad \toraxi = \pm u\,.
\end{equation}
Here $\mathcal{H}=a'(t)/a(t)$
is the Hubble function in conformal time gauge and primes denote differentiation with respect to the conformal time .

The torsion scalar for the vector branch becomes
\begin{align}
    T = \frac{1}{a^2} \left(6 \mathcal{H}^2 -12 i u \mathcal{H} - 6 u^2\right)\,,
\end{align}
while for the axial branch we find
\begin{align}
     T = \frac{1}{a^2} \left( 6\mathcal{H}^2 - 6 u^2 \right)\,,
\end{align}
which respectively govern the equations of motion for the separate branches.

Moreover, we perform a $3+1$ split of all dynamical fields into their time and space components in the next sections. For this purpose, we write the spatial metric as
\begin{equation}\label{eq:spatialtomet3D}
h_{\mu\nu}\dd x^\mu \dd x^\nu = a(t)^2\gamma_{ab}\dd x^a\dd x^b\,,
\end{equation}
where small Latin indices label spatial coordinates \(r, \vartheta, \varphi\), and the time-independent spatial metric is
\begin{equation}\label{eq:met3D}
\gamma_{ab}\dd x^a\dd x^b = \frac{\dd r^2}{1-u^2r^2}+r^2\dd\Omega^2\,.
\end{equation}
We denote its totally antisymmetric Levi-Civita tensor by \(\upsilon_{abc}\), so that
\begin{equation}\label{eq:lc3D}
\upsilon_{r\vartheta\varphi} = \frac{r^2\sin\vartheta}{\sqrt{1-u^2 r^2}}\,.
\end{equation}
Finally, we will denote the covariant derivative of the Levi-Civita connection of \(\gamma_{ab}\) by \(\dd_a\) and by $\triangle$ the corresponding Laplacian.

\subsection{Tetrad and energy-momentum perturbations}

In the following, we will consider a perturbed tetrad of the form
\begin{equation}\label{eq:cosmoperttet}
e^A{}_{\mu} = \bar{e}^A{}_{\mu} + \bar{e}^A{}_{\nu}\bar{g}^{\nu\rho}\tau_{\rho\mu}\,,
\end{equation}
where a bar denotes the unperturbed, cosmologically symmetric geometry detailed in the previous section, and the components $\tau_{\mu\nu}$ contain the perturbative degrees of freedom. We first employ the $3+1$ decomposition with help of the tensors \eqref{eq:met3D} and \eqref{eq:lc3D}. It defines the following quantities, which turn out to be very convenient to perform and present the perturbative analysis in a clear and well-readable way in Section~\ref{sec:cosmo_perturb}
\begin{gather}
\hat{\tau}_{00} = a^{-2}\tau_{00} = \hat{\phi}\,, \quad
\hat{\tau}_{0b} = a^{-2}\tau_{0b} = \dd_b\hat{j} + \hat{b}_b\,, \quad
\hat{\tau}_{a0} = a^{-2}\tau_{a0} = \dd_a\hat{y} + \hat{v}_a\,,\nonumber\\
\hat{\tau}_{ab} = a^{-2}\tau_{ab} = \hat{\psi}\gamma_{ab} + \dd_a\dd_b\hat{\sigma} + \dd_b\hat{c}_a + \upsilon_{abc}(\dd^c\hat{\xi} + \hat{w}^c) + \frac{1}{2}\hat{q}_{ab}\,.\label{eq:cosmopertdec}
\end{gather}
We see that the degrees of freedom of the field are organized in five scalars \(\hat{\phi}, \hat{j}, \hat{y}, \hat{\psi}, \hat{\sigma}\), one pseudoscalar \(\hat{\xi}\), three divergence-free vectors \(\hat{b}_a, \hat{v}_a, \hat{c}_a\), one divergence-free pseudovector \(\hat{w}_a\) and one trace-free, divergence-free symmetric tensor \(\hat{q}_{ab}\). In other words, these quantities are subject to the conditions
\begin{equation}
\dd_a\hat{b}^a = \dd_a\hat{v}^a = \dd_a\hat{c}^a = \dd_a\hat{w}^a = 0\,, \quad
\dd_a\hat{q}^{ab} = 0\,, \quad
\hat{q}_{[ab]} = 0\,, \quad
\hat{q}_a{}^a = 0\,,
\end{equation}
and thus represent all 16 degrees of freedom of the original perturbation $\tau_{\mu\nu}$.

Second, we need to introduce the energy-momentum tensor perturbations in a similar fashion to construct the field equations of the system. Conventionally, the linearly perturbations of the Hilbert energy-momentum tensor (the symmetric part of the teleparallel energy momentum tensor with lowered indices) are expanded in the form

\begin{subequations}\label{eq:cosmopertem2}
	\begin{align}
	\Theta_{00} &= a^2(\bar{\rho} + \hat{\mathcal{E}} - 2\bar{\rho}\hat{\tau}_{00})\,,\\
	\Theta_{0a} &= -a^2\left[2\bar{\rho}\hat{\tau}_{(0a)} + (\bar{\rho} + \bar{p})(\dd_a\hat{\mathcal{L}} + \hat{\mathcal{X}}_a)\right]\,,\\
	\Theta_{ab} &= a^2\left(\bar{p}\gamma_{ab} + 2\bar{p}\hat{\tau}_{(ab)} + \hat{\mathcal{P}}\gamma_{ab} + \dd_a\dd_b\hat{\mathcal{S}} - \frac{1}{3}\triangle\hat{\mathcal{S}}\gamma_{ab} + \dd_{(a}\hat{\mathcal{V}}_{b)} + \hat{\mathcal{T}}_{ab}\right)\,.
	\end{align}
\end{subequations}
The matter content is now described by the four scalars $\hat{\mathcal{E}}$, $\hat{\mathcal{P}}$, $\hat{\mathcal{L}}$, $\mathcal{S}$, the two divergence-free vectors $\hat{\mathcal{X}}_a$, $\hat{\mathcal{V}}_{b}$ and the trace and divergence-free tensor $\hat{\mathcal{T}}_{ab}$. The following combinations are interpreted as velocity perturbation
\begin{equation}
    \hat{\mathcal{U}}_a = \dd_a\hat{\mathcal{L}} + \hat{\mathcal{X}}_a\,,
\end{equation}
and anisotropic pressure perturbation
\begin{equation}
\hat{\pi}_{ab} = \dd_a\dd_b\hat{\mathcal{S}} - \frac{1}{3}\triangle\hat{\mathcal{S}}\gamma_{ab} + \dd_{(a}\hat{\mathcal{V}}_{b)} + \hat{\mathcal{T}}_{ab}\,.
\end{equation}

Finally, in the tetrad formulation, we define the perturbations of the full energy momentum tensor as
\begin{equation}
\Theta_A{}^{\mu} = \bar{\Theta}_A{}^{\mu} + \hat{\mathfrak{T}}_A{}^{\mu}\,,
\end{equation}
so that, after transforming indices with the background tetrad, we find for the perturbations of the Hilbert energy-momentum
\begin{equation}
    \Theta_{\mu\nu} - \bar{\Theta}_{\mu\nu} = \hat{\mathfrak{T}}_{\mu\nu} + \bar{g}^{\rho\sigma}(2 \tau_{\rho(\mu}\bar{\Theta}_{\nu)\sigma} + \tau_{\nu\rho}\bar{\Theta}_{\sigma\mu})\,.
\end{equation}
It allows us to express the lower index perturbation tensor $\mathfrak{T}_{\mu\nu}$ in terms of the quantities introduced in \eqref{eq:cosmopertdec} and \eqref{eq:cosmopertem2} as
\begin{subequations}\label{eq:cosmopertemcomp}
	\begin{align}
	\hat{\mathfrak{T}}_{00} &= \hat{\mathcal{E}} + \bar{\rho}\hat{\phi}\,,\\
	\hat{\mathfrak{T}}_{0b} &= -\left[(\bar{\rho} + \bar{p})\hat{\mathcal{U}}_b + \bar{p}(\hat{v}_b + \dd_b\hat{y})\right]\,,\\
	\hat{\mathfrak{T}}_{a0} &= -\left[(\bar{\rho} + \bar{p})(\hat{\mathcal{U}}_a + \hat{v}_a + \dd_a\hat{y}) + \bar{p}(\hat{b}_a + \dd_a\hat{j})\right]\,,\\
	\hat{\mathfrak{T}}_{ab} &= \hat{\mathcal{P}}\gamma_{ab} + \hat{\pi}_{ab} - \bar{p}\left[\hat{\psi}\gamma_{ab} + \dd_b\dd_a\hat{\sigma} + \dd_a\hat{c}_b - \upsilon_{abc}(\dd^c\hat{\xi} + \hat{w}^c) + \frac{1}{2}\hat{q}_{ab}\right]\,,
	\end{align}
\end{subequations}
where \(\hat{\mathcal{U}}_a\) and \(\hat{\pi}_{ab}\) are further decomposed as given above.

\subsection{Perturbed field equations}
Along a similar vein, we again employ the $3+1$ decomposition with help of the tensors \eqref{eq:met3D} and \eqref{eq:lc3D} to define another set of quantities which will be used to simplify the calculations that follow for the perturbed field equations
\begin{gather}
\hat{\mathfrak{E}}_{00} = a^{-2}\mathfrak{E}_{00} = \hat{\Phi}\,, \quad
\hat{\mathfrak{E}}_{0b} = a^{-2}\mathfrak{E}_{0b} = \dd_b\hat{J} + \hat{B}_b\,, \quad
\hat{\mathfrak{E}}_{a0} = a^{-2}\mathfrak{E}_{a0} = \dd_a\hat{Y} + \hat{V}_a\,,\nonumber\\
\hat{\mathfrak{E}}_{ab} = a^{-2}\mathfrak{E}_{ab} = \hat{\Psi}\gamma_{ab} + \dd_a\dd_b\hat{\Sigma} + \dd_a\hat{C}_b + \upsilon_{abc}(\dd^c\hat{\Xi} + \hat{W}^c) + \frac{1}{2}\hat{Q}_{ab}\,.\label{eq:cosmofeqdec}
\end{gather}
Here, in analogy to the irreducible components of \(\tau_{\mu\nu}\), the five expressions \(\hat{\Phi}, \hat{J}, \hat{Y}, \hat{\Psi}, \hat{\Sigma}\) are scalars, \(\hat{\Xi}\) is a pseudoscalar, \(\hat{B}_a, \hat{V}_a, \hat{C}_a\) are three divergence-free vectors, \(\hat{W}_a\) is a divergence-free pseudovector and \(\hat{Q}_{ab}\) is a trace-free, divergence-free, symmetric tensor. Hence, they are subject to the conditions
\begin{equation}
\dd_a\hat{B}^a = \dd_a\hat{V}^a = \dd_a\hat{C}^a = \dd_a\hat{W}^a = 0\,, \quad
\dd_a\hat{Q}^{ab} = 0\,, \quad
\hat{Q}_{[ab]} = 0\,, \quad
\hat{Q}_a{}^a = 0\,.
\end{equation}

The perturbed gravitational field equations read
\begin{equation}\label{eq:cosmopertfeq}
\bar{E}_A{}^{\mu} + \mathfrak{E}_A{}^{\mu} = E_A{}^{\mu} =  \Theta_A{}^{\mu} = \bar{\Theta}_A{}^{\mu} + \hat{\mathfrak{T}}_A{}^{\mu}\,,
\end{equation}
or equivalently their lower case spacetime index version (see also \eqref{eq:ftEOM}),
\begin{align}
    \bar E_{\mu\nu} +  \mathfrak{E}_{\mu\nu} + \bar{g}^{\rho\sigma}(2 \tau_{\rho(\mu}\bar{E}_{\nu)\sigma} + \tau_{\nu\rho}\bar{E}_{\sigma\mu}) = E_{\mu\nu} =  \Theta_{\mu\nu} = \bar{\Theta}_{\mu\nu} + \hat{\mathfrak{T}}_{\mu\nu} + \bar{g}^{\rho\sigma}(2 \tau_{\rho(\mu}\bar{\Theta}_{\nu)\sigma} + \tau_{\nu\rho}\bar{\Theta}_{\sigma\mu})\,.
\end{align}
The background geometry part of the field equations $\bar E_{\mu\nu}$ can be decomposed into
\begin{align}\label{eq:bgeq}
    \bar E_{\mu\nu} = \mathfrak{N}n_\mu n_\nu + \mathfrak{H}h_{\mu\nu}\,,
\end{align}
where the normal covector $n$ and the spatial metric have been introduced in \eqref{eq:nh}. Comparing this to the decomposition \eqref{eq:cosmopertem2} of the energy momentum tensor yields that the background field equations reduce to $\mathfrak{N} = \bar\rho$ and $\mathfrak{H} = \bar p$. If the background equations are satisfied, the field equations reduce to the perturbation equations
\begin{equation}
\mathfrak{E}_{\mu\nu} = \hat{\mathfrak{T}}_{\mu\nu}\,.
\end{equation}
In total we can now use the $3+1$ decomposition of the tetrad \eqref{eq:cosmopertdec}, the energy-momentum tensor \eqref{eq:cosmopertem2} and the field equations \eqref{eq:cosmofeqdec} to obtain
\begin{itemize}
    \item six scalar equations
        \begin{gather}
        \hat{J} = -(\bar{\rho} + \bar{p})\hat{\mathcal{L}} - \bar{p}\hat{y}\,, \quad
        \hat{\Sigma} = \hat{\mathcal{S}} + \bar{p}\hat{\sigma}\,, \quad
        \hat{\Xi} = \bar{p}\hat{\xi}\,,\nonumber\\
        \hat{\Psi} = \hat{\mathcal{P}} - \frac{1}{3}\triangle\hat{\mathcal{S}} - \bar{p}\hat{\psi}\,, \quad
        \hat{\Phi} = \hat{\mathcal{E}} + \bar{\rho}\hat{\phi}\,, \quad
        \hat{Y} = -(\bar{\rho} + \bar{p})(\hat{\mathcal{L}} + \hat{y}) - \bar{p}\hat{j}\,,\label{eq:pertscalgen}
        \end{gather}
    \item four vector equations
        \begin{gather}\
        \hat{V}_a = -(\bar{\rho} + \bar{p})(\hat{\mathcal{X}}_a + \hat{v}_a) - \bar{p}\hat{b}_a\,, \quad
        \hat{C}_a = \hat{\mathcal{V}}_a - \bar{p}\hat{c}_a\,,\nonumber\\
        \hat{W}_a = \bar{p}\hat{w}_a - \frac{1}{2}\upsilon_{abc}\dd^b\hat{\mathcal{V}}^c\,, \quad
        \hat{B}_a = -(\bar{\rho} + \bar{p})\hat{\mathcal{X}}_b - \bar{p}\hat{v}_b\,,\label{eq:pertvecgen}
        \end{gather}
    \item and a tensor equation
        \begin{equation}\label{eq:perttengen}
        \hat{Q}_{ab} = 2\hat{\mathcal{T}}_{ab} - \bar{p}\hat{q}_{ab}\,.
        \end{equation}
\end{itemize}

Together, these represent the entire system of perturbed equations for teleparallel gravity cosmology.

\subsection{Gauge transformations and gauge invariant quantities}

The tetrad~\eqref{eq:cosmoperttet} retains its form as a small perturbation around the cosmologically symmetric background tetrad under an infinitesimal coordinate transformation
\begin{equation}\label{eq:lincoordtrans}
x'^{\mu} = x^{\mu} + X^{\mu}\,.
\end{equation}
Under this transformation, the tetrad perturbation changes to
\begin{equation}
\tau_{\mu\nu}' = \tau_{\mu\nu} - \bar{\nabla}_{\nu}X_{\mu} + \bar{T}_{\mu\nu}{}^{\rho}X_{\rho} = \tau_{\mu\nu} - \lc{\bar{\nabla}}_{\nu}X_{\mu} - \bar{K}_{\mu\nu}{}^{\rho}X_{\rho}\,.
\end{equation}
By making a suitable choice of this transformation, and decomposing the transformation as
\begin{equation}
\hat{X}_0 = a^{-1}X_0 = \hat{X}_{\perp}\,, \quad
\hat{X}_a = a^{-1}X_a = \dd_a\hat{X}_{\parallel} + \hat{Z}_a
\end{equation}
in analogy to the irreducible decomposition of the perturbed geometry, see \eqref{eq:cosmopertdec}, one can eliminate certain components in the irreducible decomposition of the perturbations. We will denote such a fixed choice, or gauge, with a letter, e.g., \(\mathsf{G}\), under the corresponding quantity, and write these gauge-fixed quantities with boldface letters. Instead of gauge-fixed, one may also use the term gauge-invariant quantities, since they are independent of the gauge prior to applying the gauge transformation. Performing the irreducible decomposition of the gauge transformations, one finds the gauge-independent tetrad perturbations \eqref{eq:cosmopertdec}
\begin{gather}
\ginv{\psi}{G} = \hat{\psi} + \frac{\mathcal{H}\gdep{X}{G}_{\perp}}{a}\,, \quad
\ginv{\sigma}{G} = \hat{\sigma} - \frac{\gdep{X}{G}_{\parallel}}{a}\,, \quad
\ginv{y}{G} = \hat{y} - \frac{\gdep{X}{G}_{\parallel}' - \torvec\gdep{X}{G}_{\parallel}}{a}\,,\nonumber\\
\ginv{j}{G} = \hat{j} - \frac{\gdep{X}{G}_{\perp} + (\torvec - \mathcal{H})\gdep{X}{G}_{\parallel}}{a}\,, \quad
\ginv{\xi}{G} = \hat{\xi} + \frac{\toraxi\gdep{X}{G}_{\parallel}}{a}\,, \quad
\ginv{\phi}{G} = \hat{\phi} - \frac{\gdep{X}{G}_{\perp}'}{a}\,,\nonumber\\
\ginv{c}{G}_a = \hat{c}_a - \frac{\gdep{Z}{G}_a}{a}\,, \quad
\ginv{v}{G}_a = \hat{v}_a - \frac{\gdep{Z}{G}_a' - \torvec\gdep{Z}{G}_a}{a}\,, \quad
\ginv{b}{G}_a = \hat{b}_a - \frac{(\torvec - \mathcal{H})\gdep{Z}{G}_a}{a}\,, \quad
\ginv{w}{G}_a = \hat{w}_a + \frac{\toraxi\gdep{Z}{G}_a}{a}\,, \quad
\ginv{q}{G} = \hat{q}_{ab}\,,\label{eq:gtrandecct}
\end{gather}
the energy-momentum perturbations \eqref{eq:cosmopertem2}
\begin{gather}
\ginv{\mathcal{E}}{G} = \hat{\mathcal{E}} + a^{-1}\gdep{X}{G}_{\perp}\bar{\rho}'\,, \quad
\ginv{\mathcal{P}}{G} = \hat{\mathcal{P}} + a^{-1}\gdep{X}{G}_{\perp}\bar{p}'\,, \quad
\ginv{\mathcal{L}}{G} = \hat{\mathcal{L}} + (a^{-1}\gdep{X}{G}_{\parallel})'\,, \quad
\ginv{\mathcal{X}}{G}_a = \hat{\mathcal{X}}_a + (a^{-1}\gdep{Z}{G}_a)'\,,\nonumber\\
\ginv{\mathcal{S}}{G} = \hat{\mathcal{S}}\,, \quad
\ginv{\mathcal{V}}{G}_a = \hat{\mathcal{V}}_a\,, \quad
\ginv{\mathcal{T}}{G}_{ab} = \hat{\mathcal{T}}_{ab}\,,
\end{gather}
as well as the field equation components \eqref{eq:pertscalgen}, \eqref{eq:pertvecgen} and \eqref{eq:perttengen},
\begin{gather}
\ginv{V}{G}_a = \hat{V}_a - \frac{(\torvec - \mathcal{H})\mathfrak{N}\gdep{Z}{G}_a}{a}\,, \quad
\ginv{\Phi}{G} = \hat{\Phi} - \frac{\mathfrak{N}\gdep{X}{G}_{\perp}' - \gdep{X}{G}_{\perp}\mathfrak{N}'}{a}\,, \quad
\ginv{\Psi}{G} = \hat{\Psi} - \frac{(\mathfrak{H}\mathcal{H} - \mathfrak{H}')\gdep{X}{G}_{\perp}}{a}\,, \quad
\ginv{\Sigma}{G} = \hat{\Sigma} + \frac{\mathfrak{H}\gdep{X}{G}_{\parallel}}{a}\,,\nonumber\\
\ginv{\Xi}{G} = \hat{\Xi} + \frac{\mathfrak{H}\toraxi\gdep{X}{G}_{\parallel}}{a}\,, \quad
\ginv{J}{G} = \hat{J} - \frac{[(\torvec - \mathcal{H})\mathfrak{H} - \mathcal{H}\mathfrak{N}]\gdep{X}{G}_{\parallel} + \mathfrak{N}\gdep{X}{G}_{\parallel}'}{a}\,, \quad
\ginv{Y}{G} = \hat{Y} - \frac{(\torvec - \mathcal{H})\mathfrak{N}\gdep{X}{G}_{\parallel} - \mathfrak{H}\gdep{X}{G}_{\perp}}{a}\,,\nonumber\\
\ginv{C}{G}_a = \hat{C}_a + \frac{\mathfrak{H}\gdep{Z}{G}_a}{a}\,, \quad
\ginv{W}{G}_a = \hat{W}_a + \frac{\mathfrak{H}\toraxi\gdep{Z}{G}_a}{a}\,, \quad
\ginv{B}{G}_a = \hat{B}_a - \frac{[(\torvec - \mathcal{H})\mathfrak{H} - \mathcal{H}\mathfrak{N}]\gdep{Z}{G}_a + \mathfrak{N}\gdep{Z}{G}_a'}{a}\,, \quad
\ginv{Q}{G}_{ab} = \hat{Q}_{ab}\,.
\end{gather}

A fixed choice of a gauge can be specified in two possible ways: either by imposing conditions on the gauge fixed perturbations appearing on the left-hand side of any of the equations listed above, or by expressing the gauge transformation \(\gdep{X}{G}\), which is necessary in order to transform the perturbations from an arbitrary to the desired gauge, in terms of these arbitrary-gauge perturbations. Here we give both specifications for each of the gauges we will use in the remainder of this article. The ``zero gauge'' \(\mathsf{G} = \mathsf{0}\) used to construct the gauge-invariant quantities in~\cite{Hohmann:2020vcv} is obtained by the gauge conditions
\begin{equation}
\ginv{\sigma}{0} = \ginv{j}{0} = 0\,, \quad \ginv{c}{0}_a = 0\,,
\end{equation}
which are satisfied if the gauge transformation from an arbitrary gauge is chosen as
\begin{equation}
a^{-1}\gdep{X}{0}_{\perp} = \hat{j} + (\mathcal{H} - \torvec)\hat{\sigma}\,, \quad
a^{-1}\gdep{X}{0}_{\parallel} = \hat{\sigma}\,, \quad
a^{-1}\gdep{Z}{0}_a = \hat{c}_a\,.
\end{equation}
In~\cite{Golovnev:2018wbh} the Newtonian gauge \(\mathsf{G} = \mathsf{N}\) is used, where \(\ginv{j}{N} = -\ginv{y}{N}\), and so \(\ginv{\psi}{N} = \ginv{\psi}{0} + \mathcal{H}\ginv{y}{0}\) and \(\ginv{\phi}{N} = \ginv{\phi}{0} - \mathcal{H}\ginv{y}{0} - \ginv{y}{0}'\). The gauge conditions read
\begin{equation}
\ginv{j}{N} + \ginv{y}{N} = \ginv{\sigma}{N} = 0\,, \quad \ginv{b}{N}_a + \ginv{v}{N}_a = 0\,.
\end{equation}
For the vector perturbations, the gauge transformation needed to satisfy these conditions is determined only up to a contribution constant in time, and one has
\begin{equation}
a^{-1}\gdep{X}{N}_{\perp} = \hat{j} + \hat{y} - \hat{\sigma}'\,, \quad
a^{-1}\gdep{X}{N}_{\parallel} = \hat{\sigma}\,, \quad
(a^{-1}\gdep{Z}{N}_a)' = \hat{b}_a + \hat{v}_a\,.
\end{equation}
For the fluid matter, also the comoving gauge \(\mathsf{G} = \mathsf{C}\) is used, which is defined by the conditions
\begin{equation}
\ginv{j}{C} + \ginv{y}{C} = \ginv{\mathcal{L}}{C} = 0\,, \quad \ginv{\mathcal{X}}{C}_a = 0\,.
\end{equation}
In this case the gauge transformation reads
\begin{equation}
a^{-1}\gdep{X}{C}_{\perp} = \hat{j} + \hat{y} + \hat{\mathcal{L}}\,, \quad
(a^{-1}\gdep{X}{C}_{\parallel})' = -\hat{\mathcal{L}}\,, \quad
(a^{-1}\gdep{Z}{C}_a)' = -\hat{\mathcal{X}}_a\,.
\end{equation}
In the following sections, it shall be clear from the notation which gauge is used in the definition of the appearing quantities.

\subsection{Harmonic expansion}\label{ssec:harmonic}

The last ingredient we need to analyze the dynamics of the cosmological perturbations in all detail in Section \ref{sec:cosmo_perturb}, is the convenient expansion of the perturbations into a harmonic basis composed from eigenfunctions of the Laplace operator \(\triangle = \dd^a\dd_a\) of the spatial background model space. For the cosmological FLRW models, where these spaces are maximally symmetric, three-dimensional, Riemannian manifolds, these harmonic tensors have been discussed extensively in the literature~\cite{Cohl2012,grellier:hal-00022000,Sandberg,Abbott:1986ct}. For the scalar perturbations, one finds that there exist harmonics \(\mathfrak{s}(\beta)\) satisfying
\begin{equation}
\triangle\mathfrak{s}(\beta) = -k^2\mathfrak{s}(\beta) = -(\beta^2 - u^2)\mathfrak{s}(\beta)\,,
\end{equation}
where \(\beta \in \{3, 4, 5, \ldots\}\) for \(u^2 = 1\) and \(\beta \geq 0\) for $u^2\in\{-1,0\}$~\cite{Abbott:1986ct}. We then continue with the divergence-free vectors. It is convenient to introduce the curl
\begin{equation}
\curl\hat{z}_a = \upsilon_{abc}\dd^b\hat{z}^c\,.
\end{equation}
It is easy to check that
\begin{equation}
\curl\curl\hat{z}_a = -\triangle\hat{z}_a + 2u^2\hat{z}_a\,.
\end{equation}
One finds that the harmonics are given by the two helicities \(\hat{\mathfrak{v}}_a^{\pm}(\beta)\) satisfying
\begin{equation}
\curl\hat{\mathfrak{v}}_a^{\pm}(\beta) = \pm\beta\hat{\mathfrak{v}}_a^{\pm}(\beta)\,,
\end{equation}
where \(\beta\) takes the same values as before. It follows that they are eigenfunctions of the Laplacian with
\begin{equation}
k^2\hat{\mathfrak{v}}_a^{\pm}(\beta) = -\triangle\hat{\mathfrak{v}}_a^{\pm}(\beta) = (\beta^2 - 2u^2)\hat{\mathfrak{v}}_a^{\pm}(\beta)\,.
\end{equation}
Finally, for the symmetric, trace-free, divergence-free tensors one can similarly introduce a curl
\begin{equation}
\curl\hat{z}_{ab} = \upsilon_{cd(a}\dd^c\hat{z}_{b)}{}^d\,,
\end{equation}
which now satisfies
\begin{equation}
\curl\curl\hat{z}_{ab} = -\triangle\hat{z}_{ab} + 3u^2\hat{z}_{ab}\,.
\end{equation}
Again the harmonics come in two helicities \(\hat{\mathfrak{t}}_{ab}^{\pm}(\beta)\) satisfying
\begin{equation}
\curl\hat{\mathfrak{t}}_{ab}^{\pm}(\beta) = \pm\beta\hat{\mathfrak{t}}_{ab}^{\pm}(\beta)\,,
\end{equation}
where \(\beta\) takes again the same values as in the scalar case. It follows that they are eigenfunctions of the Laplacian with
\begin{equation}
k^2\hat{\mathfrak{t}}_{ab}^{\pm}(\beta) = -\triangle\hat{\mathfrak{t}}_{ab}^{\pm}(\beta) = (\beta^2 - 3u^2)\hat{\mathfrak{t}}_{ab}^{\pm}(\beta)\,.
\end{equation}
We will make use of the allowed ranges of \(k^2\) when we derive the perturbed field equations.

\section{Cosmological perturbations: perturbed equations in \texorpdfstring{$f(T)$}{f(T)} gravity}\label{sec:cosmo_perturb}
In this Section we evaluate the field equations \eqref{eq:pertscalgen}, \eqref{eq:pertvecgen} and \eqref{eq:perttengen} for the perturbations in general $f(T)$-gravity models. In particular we analyse the influence of the curvature parameter $u^2$ on the number of degrees of freedom and their propagation behaviour. We find that in the flat case $u^2=0$, one of the perturbations is not determined, while in the curved case $u^2\neq 0$ all perturbations are determined by the field equations.

\subsection{Background equations}
We start our discussion of the cosmological dynamics of $f(T)$ gravity by a brief review of its background field equations, which we display here using the conformal time coordinate; see~\cite{Hohmann:2020zre} for their form in cosmological time. For the vector branch~\eqref{eq:vectet}, they take the form
\begin{subequations}
	\begin{align}
	f - 12\frac{f_T}{a^2}\mathcal{H}(\mathcal{H} + iu) &= 2\kappa^2\bar{\rho}\,,\\
	-f + 4\frac{f_T}{a^2}(2\mathcal{H}^2 + 3iu\mathcal{H} - u^2 + \mathcal{H}') + 48\frac{f_{TT}}{a^4}(\mathcal{H} + iu)^2[\mathcal{H}' - \mathcal{H}(\mathcal{H} + iu)] &= 2\kappa^2\bar{p}\,,
	\end{align}\label{eq:back_FE_vector}
\end{subequations}
while for the axial tetrad~\eqref{eq:axtet} they read
\begin{subequations}
	\begin{align}
	f - 12\frac{f_T}{a^2}\mathcal{H}^2 &= 2\kappa^2\bar{\rho}\,,\\
	-f + 4\frac{f_T}{a^2}(2\mathcal{H}^2 - u^2 + \mathcal{H}') + 48\frac{f_{TT}}{a^4}\mathcal{H}^2(\mathcal{H}' + u^2 - \mathcal{H}^2) &= 2\kappa^2\bar{p}\,.
	\end{align}\label{eq:back_FE_axial}
\end{subequations}
In the flat limiting case \(u \to 0\), both branches converge: the scalar functions in the torsion tensor~\eqref{eq:cosmotor} take the common value
\begin{equation}
\torvec = \mathcal{H}\,, \quad
\toraxi = 0\,.
\end{equation}
and the background field equations clearly assume the form
\begin{subequations}
	\begin{align}
	f - 12\frac{f_T}{a^2}\mathcal{H}^2 &= 2\kappa^2\bar{\rho}\,,\\
	-f + 4\frac{f_T}{a^2}(2\mathcal{H}^2 + \mathcal{H}') + 48\frac{f_{TT}}{a^4}\mathcal{H}^2(\mathcal{H}' - \mathcal{H}^2) &= 2\kappa^2\bar{p}\,.
	\end{align}
\end{subequations}
In the following, we will assume that the background equations are satisfied, so that we can freely exchange the background values \(\bar{\rho}\) and \(\bar{p}\) of the matter density and pressure by the corresponding geometry sides of the background field equations and vice versa.

\subsection{Tensorial perturbations}\label{ssec:tensorpert}
We start our analysis of the cosmological perturbations with the tensor sector. Recall that the tensor field equation \eqref{eq:perttengen} is given by
\begin{equation}
\frac{1}{2}\hat{Q}_{ab} = \hat{\mathcal{T}}_{ab} - \frac{1}{2}\bar{p}\hat{q}_{ab}\,,
\end{equation}
which we will now evaluate for the general \(f(T)\) class of gravity theories for the different branches of cosmological backgrounds.

\subsubsection{Vector branch}
For the vector branch, see~\eqref{eq:vectet} and~\eqref{eq:vectorbranch}, we find that the field equation for the tensor perturbations takes the form
\begin{equation} \label{eq:gwpe_vec}
2\kappa^2a^2\ginv{\mathcal{T}}{0}_{ab} = f_T\left(\triangle\ginv{q}{0}_{ab} - 2u^2\ginv{q}{0}_{ab} - 2\mathcal{H}\ginv{q}{0}_{ab}' - \ginv{q}{0}_{ab}''\right) + 12\frac{f_{TT}}{a^2}(\mathcal{H} + iu)[\mathcal{H}(\mathcal{H} + iu) - \mathcal{H}']\left(\ginv{q}{0}_{ab}' - iu\ginv{q}{0}_{ab}\right)\,.
\end{equation}
Note that despite the appearance of the imaginary unit \(i\), this equation is real, since \(u\) is imaginary for the vector branch, \(u^2 < 0\). Note that the \(f_T\) term is simply the usual wave equation on a spatially curved FLRW background, while the \(f_{TT}\) term constitutes a modification to the Hubble friction and curvature terms only. This means that the speed of gravitational wave is equal to the speed of light.

\subsubsection{Axial branch}
For the axial branch, see~\eqref{eq:axtet} and~\eqref{eq:axialbranch}, the tensor perturbations are governed by the equation
\begin{equation} \label{eq:gwpe_ax}
2\kappa^2a^2\ginv{\mathcal{T}}{0}_{ab} = f_T\left(\triangle\ginv{q}{0}_{ab} - 2u^2\ginv{q}{0}_{ab} - 2\mathcal{H}\ginv{q}{0}_{ab}' - \ginv{q}{0}_{ab}''\right) + 12\frac{f_{TT}}{a^2}\mathcal{H}(\mathcal{H}^2 - u^2 - \mathcal{H}')\ginv{q}{0}_{ab}'\,.
\end{equation}
The qualitative structure of this equation is similar to the vector branch, with the \(f_T\) term resembling the usual wave equation, while the \(f_{TT}\) term contributes to the Hubble friction only. As in the vector branch, the speed of gravitational waves is exactly the same as the speed of light.

\subsubsection{Flat case}
In the limit \(u \to 0\), the previously shown equations reduce to the flat case
\begin{equation}
2\kappa^2a^2\ginv{\mathcal{T}}{0}_{ab} = f_T\left(\triangle\ginv{q}{0}_{ab} - 2\mathcal{H}\ginv{q}{0}_{ab}' - \ginv{q}{0}_{ab}''\right) + 12\frac{f_{TT}}{a^2}\mathcal{H}(\mathcal{H}^2 - \mathcal{H}')\ginv{q}{0}_{ab}'\,.
\end{equation}
This result agrees with what has been found in~\cite{Golovnev:2018wbh}. In all the three cases the tensorial perturbations are determined by the perturbative field equations.

\subsection{Vectorial perturbations}\label{ssec:vec_pert}
For the vectorial perturbations, we found four field equations~\eqref{eq:pertvecgen}. One of these equations becomes redundant; on the gravitational side of the field equations it corresponds to a Bianchi identity. On the energy-momentum side, it corresponds to the vector part of the energy-momentum conservation. Defining the gauge-invariant variable
\begin{equation}
\hat{\mathcal{Q}}_a = (\bar{\rho} + \bar{p})(\hat{\mathcal{X}}_a + \hat{v}_a + \hat{b}_a)\,,
\end{equation}
this equation takes the form
\begin{equation}
\hat{\mathcal{Q}}_a' + 4\mathcal{H}\hat{\mathcal{Q}}_a + \frac{1}{2}\triangle\hat{\mathcal{V}}_a + u^2\hat{\mathcal{V}}_a = 0\,.
\end{equation}
The remaining, independent equations are the spatial and mixed part of the antisymmetric field equations as well as the mixed part of the symmetric equations,
\begin{equation}
E_{[ab]} = \Theta_{[ab]} = 0\,, \quad
E_{[a0]} = \Theta_{[a0]} = 0\,, \quad
E_{(a0)} = \Theta_{(a0)} = 0\,.
\end{equation}
These equations decompose into the irreducible components which are combinations of the equations \eqref{eq:pertvecgen}
\begin{subequations}
\begin{align}
\dd_{[a}\hat{C}_{b]} + \bar{p}\dd_{[a}\hat{c}_{b]} + 2\upsilon_{abc}(\hat{W}^c - \bar{p}\hat{w}^c) &= 0\,,\\
\hat{V}_a - \hat{B}_a + \bar{\rho}\hat{v}_a + \bar{p}\hat{b}_a &= 0\,,\\
\hat{V}_a + \hat{B}_a + (\bar{\rho} + 2\bar{p})\hat{v}_a + \bar{p}\hat{b}_a + 2(\bar{\rho} + \bar{p})\hat{\mathcal{X}}_a &= 0\,.
\end{align}
\end{subequations}
In the following we will analyze these equations using the zero gauge condition \(\ginv{c}{0}_a = 0\), starting with the antisymmetric equations, before we continue towards the symmetric equations.

\subsubsection{Vector branch}
We start again with the vector branch (\eqref{eq:vectet},~\eqref{eq:vectorbranch}). In this case the two antisymmetric equations read
\begin{subequations}
\begin{align}
f_{TT}(\mathcal{H} + iu)[\mathcal{H}(\mathcal{H} + iu) - \mathcal{H}'](\dd_{[a}\ginv{b}{0}_{b]} + iu\upsilon_{abc}\ginv{w}{0}^c) &= 0\,,\\
f_{TT}(\mathcal{H} + iu)[\mathcal{H}(\mathcal{H} + iu) - \mathcal{H}'](\upsilon_{abc}\dd^b\ginv{w}{0}^c - 2iu\ginv{b}{0}_a) &= 0\,.
\end{align}
\end{subequations}
For a non-trivial theory we assume \(f_{TT} \neq 0\), and also the remaining terms in brackets are non-vanishing. This yields two coupled equations for \(\ginv{b}{0}_a\) and \(\ginv{w}{0}_a\). In order to decouple these equations, one can take the curl of both of them. Omitting the factors in front of the equations, the result reads
\begin{subequations}\label{eq:vecvec1}
\begin{align}
\triangle\ginv{w}{0}_a + 2iu\upsilon_{abc}\dd^b\ginv{b}{0}^c - 2u^2\ginv{w}{0}_a &= 0\,,\\
\triangle\ginv{b}{0}_a + 2iu\upsilon_{abc}\dd^b\ginv{w}{0}^c - 2u^2\ginv{b}{0}_a &= 0\,.
\end{align}
\end{subequations}
The terms involving the Levi-Civita tensor \(\upsilon_{abc}\) can now be eliminated by substituting the original equations. This finally yields the two decoupled Helmholtz equations
\begin{equation}\label{eq:vecvec2}
\triangle\ginv{w}{0}_a - 6u^2\ginv{w}{0}_a = \triangle\ginv{b}{0}_a - 6u^2\ginv{b}{0}_a = 0\,.
\end{equation}
Performing the harmonic expansion shown in Sec.~\ref{ssec:harmonic}, one now sees that this is solved only by the mode with wavenumber \(k^2 = -6u^2\), and hence \(\beta^2 = -4u^2\). Recalling \(u^2 < 0\) for the vector branch, there exists a single solution. Due to the absence of sources, the solution to \eqref{eq:vecvec2} is given by $\ginv{w}{0}_a=\ginv{b}{0}_a=0$.

The remaining perturbation \(\ginv{v}{0}_a\) is determined from the last independent equation, which is the vector part of the mixed symmetric equation. In terms of the perturbation variables, this equation reads
\begin{multline}\label{eq:vecvec3}
a^2f_T\left[\triangle(\ginv{v}{0}_a + \ginv{b}{0}_a) - 2(2\mathcal{H}^2 + u^2 - 2\mathcal{H}')(\ginv{v}{0}_a + \ginv{b}{0}_a) - 4(\mathcal{H}^2 + u^2 - \mathcal{H}')\ginv{\mathcal{X}}{0}_a\right]\\
- 12f_{TT}(\mathcal{H} + iu)[\mathcal{H}(\mathcal{H} + iu) - \mathcal{H}']\left[\upsilon_{abc}\dd^b\ginv{w}{0}^c + 4(\mathcal{H} + iu)(\ginv{\mathcal{X}}{0}_a + \ginv{v}{0}_a) + 2(2\mathcal{H} + iu)\ginv{b}{0}_a\right] = 0\,.
\end{multline}
Here we can eliminate the previously determined perturbations \(\ginv{b}{0}_a\) and \(\ginv{w}{0}_a\), which yields
\begin{equation}
a^2f_T\left[\triangle\ginv{v}{0}_a - 2(2\mathcal{H}^2 + u^2 - 2\mathcal{H}')\ginv{v}{0}_a - 4(\mathcal{H}^2 + u^2 - \mathcal{H}')\ginv{\mathcal{X}}{0}_a\right] - 48f_{TT}(\mathcal{H} + iu)^2[\mathcal{H}(\mathcal{H} + iu) - \mathcal{H}'](\ginv{\mathcal{X}}{0}_a + \ginv{v}{0}_a) = 0\,.
\end{equation}
This gives us a healthy vector branch for the non-flat FLRW background.

\subsubsection{Axial branch}
For the axial branch, (~\eqref{eq:axtet},~\eqref{eq:axialbranch}), we obtain the antisymmetric field equations
\begin{subequations}
\begin{align}
f_{TT}\mathcal{H}(\mathcal{H}^2 - u^2 - \mathcal{H}')(\dd_{[a}\ginv{b}{0}_{b]} - u\upsilon_{abc}\ginv{b}{0}^c) &= 0\,,\\
f_{TT}\mathcal{H}(\mathcal{H}^2 - u^2 - \mathcal{H}')(\upsilon_{abc}\dd^b\ginv{w}{0}^c - 2u\ginv{w}{0}_a) &= 0\,.
\end{align}
\end{subequations}
Also here we assume \(f_{TT} \neq 0\), and the remaining factors in front of the equations are non-vanishing. In this case we find that \(\ginv{b}{0}_a\) and \(\ginv{w}{0}_a\) decouple from each other, but the equations couple the polar and axial modes for each of these perturbations. To study the solutions of these equations, it is instructive to first calculate the curl of both equations. Omitting the non-vanishing factors, one then finds the equations
\begin{subequations}\label{eq:vecax1}
\begin{align}
\triangle\ginv{w}{0}_a + 2u\upsilon_{abc}\dd^b\ginv{w}{0}^c - 2u^2\ginv{w}{0}_a &= 0\,,\\
\triangle\ginv{b}{0}_a - 2u\upsilon_{abc}\dd^b\ginv{b}{0}^c - 2u^2\ginv{b}{0}_a &= 0\,.
\end{align}
\end{subequations}
As for the vector branch, one can eliminate the Levi-Civita tensor \(\upsilon_{abc}\) by substituting the original equations. This yields the Helmholtz equations
\begin{equation}\label{eq:vecax2}
\triangle\ginv{w}{0}_a + 2u^2\ginv{w}{0}_a = \triangle\ginv{b}{0}_a + 2u^2\ginv{b}{0}_a = 0\,.
\end{equation}
With the harmonic expansion from Sec.~\ref{ssec:harmonic}, this equation restricts the wavenumber to \(k^2 = 2u^2\), and hence \(\beta^2 = 4u^2\), which has no solution within the allowed range of \(\beta\). Hence, we conclude \(\ginv{b}{0}_a = \ginv{w}{0}_a = 0\). Also here the remaining perturbation \(\ginv{v}{0}_a\) is determined from the vector part of the mixed symmetric equation. In terms of the perturbation variables, this equation now reads
\begin{multline}\label{eq:vecax3}
a^2f_T\left[\triangle(\ginv{v}{0}_a + \ginv{b}{0}_a) - 2(2\mathcal{H}^2 + u^2 - 2\mathcal{H}')(\ginv{v}{0}_a + \ginv{b}{0}_a) - 4(\mathcal{H}^2 + u^2 - \mathcal{H}')\ginv{\mathcal{X}}{0}_a\right]\\
- 12f_{TT}\mathcal{H}(\mathcal{H}^2 - u^2 - \mathcal{H}')\left[\upsilon_{abc}\dd^b\ginv{w}{0}^c - 2u\ginv{w}{0}_a + 4\mathcal{H}(\ginv{\mathcal{X}}{0}_a + \ginv{v}{0}_a + \ginv{b}{0}_a)\right] = 0\,.
\end{multline}
Again we eliminate the previously determined perturbations \(\ginv{b}{0}_a\) and \(\ginv{w}{0}_a\), yielding
\begin{equation}\label{eq:vecax4}
a^2f_T\left[\triangle\ginv{v}{0}_a - 2(2\mathcal{H}^2 + u^2 - 2\mathcal{H}')\ginv{v}{0}_a - 4(\mathcal{H}^2 + u^2 - \mathcal{H}')\ginv{\mathcal{X}}{0}_a\right] - 48f_{TT}\mathcal{H}^2(\mathcal{H}^2 - u^2 - \mathcal{H}')(\ginv{\mathcal{X}}{0}_a + \ginv{v}{0}_a) = 0\,.
\end{equation}
This means that for any arbitrary $f(T)$ model, the vector branch for this background must decay for this relation to be satisfied which is inline with our expectation for the vector modes.

\subsubsection{Flat case}
Finally, we discuss the common flat limiting case of the two previously discussed spatially curved cases. In this case the antisymmetric part of the field equations yields the two equations
\begin{subequations}
\begin{align}
f_{TT}\mathcal{H}(\mathcal{H}^2 - \mathcal{H}')\dd_{[a}\ginv{b}{0}_{b]} &= 0\,,\\
f_{TT}\mathcal{H}(\mathcal{H}^2 - \mathcal{H}')\upsilon_{abc}\dd^b\ginv{w}{0}^c &= 0\,.
\end{align}
\end{subequations}
Note that the second equation is misprinted in~\cite{Golovnev:2018wbh}. By taking the curl, one now immediately obtains the Laplace equations
\begin{equation}
\triangle\ginv{b}{0}_a = \triangle\ginv{w}{0}_a = 0\,.
\end{equation}
The only solution to these equations which is compatible with the boundary conditions is spatially constant and can thus be absorbed into the homogeneous background solution. We are left with the potential \(\ginv{v}{0}_a\), which is determined by the symmetric equation
\begin{equation}
a^2f_T\triangle(\ginv{v}{0}_a + \ginv{b}{0}_a) + 4(\mathcal{H}' - \mathcal{H}^2)\left[3f_{TT}\mathcal{H}\upsilon_{abc}\dd^b\ginv{w}{0}^c + (a^2f_T + 12f_{TT}\mathcal{H}^2)(\ginv{\mathcal{X}}{0}_a + \ginv{v}{0}_a + \ginv{b}{0}_a)\right] = 0\,.
\end{equation}
Also here we can eliminate the potentials \(\ginv{b}{0}_a\) and \(\ginv{w}{0}_a\), which yields the equation
\begin{equation}
a^2f_T\triangle\ginv{v}{0}_a + 4(\mathcal{H}' - \mathcal{H}^2)(a^2f_T + 12f_{TT}\mathcal{H}^2)(\ginv{\mathcal{X}}{0}_a + \ginv{v}{0}_a) = 0\,.
\end{equation}

As with the tensorial perturbations, we found that for all the three curvature cases the vectorial perturbations are completely determined by the perturbative field equations.

\subsection{Scalar perturbations}\label{ssec:scal_pert}
Finally, we come to the scalar and pseudo-scalar perturbations. From the field equations \eqref{eq:pertscalgen} one obtains the six independent (pseudo-)scalar perturbation equations
\begin{subequations}\label{eq:pertscalsep}
\begin{align}
\hat{\Xi} - \bar{p}\hat{\xi} &= 0\,,\label{eq:pseudoscalar}\\
\hat{J} - \hat{Y} - \bar{\rho}\hat{y} - \bar{p}\hat{j} &= 0\label{eq:asymscalar}\,,\\
\hat{\Phi} - \bar{\rho}\hat{\phi} &= \hat{\mathcal{E}}\label{eq:timescalar}\,,\\
\hat{Y} + (\bar{\rho} + \bar{p})(\hat{\mathcal{L}} + \hat{y}) + \bar{p}\hat{j} &= 0\label{eq:mixedscalar}\\
\hat{\Sigma} - \bar{p}\hat{\sigma} &= \hat{\mathcal{S}}\label{eq:offdiagscalar}\,,\\
\hat{\Psi} + \triangle\hat{\Sigma} + \bar{p}(\hat{\psi} + \triangle\sigma) &= \hat{\mathcal{P}} + \frac{2}{3}\triangle\hat{\mathcal{S}}\label{eq:tracescalar}\,.
\end{align}
\end{subequations}
These are complemented by two scalar components of the Bianchi identities. In the following, we will discuss the scalar equations in the Newtonian gauge.

\subsubsection{Vector branch}\label{sssec:scalvec}
As previously, we start with the vector branch, see~\eqref{eq:vectet} and~\eqref{eq:vectorbranch}, for the cosmological background. In this case the pseudo-scalar equation~\eqref{eq:pseudoscalar}, obtained from the antisymmetric part of the spatial equations, reads
\begin{equation}\label{eq:pseudoscalarvec}
12uf_{TT}(\mathcal{H} + iu)[\mathcal{H}(\mathcal{H} + iu) - \mathcal{H}']\ginv{\xi}{N} = 0\,.
\end{equation}
We see that the pseudo-scalar \(\ginv{\xi}{N}\) decouples from the remaining perturbations and must vanish identically, \(\ginv{\xi}{N} = 0\). We continue with the antisymmetric equation with mixed indices~\eqref{eq:asymscalar}, which yields
\begin{equation}\label{eq:ypoissvecn}
f_{TT}(\mathcal{H} + iu)[3(\mathcal{H} + iu)(\mathcal{H}\ginv{\phi}{N} + \mathcal{H}\ginv{\psi}{N} - iu\ginv{\psi}{N} + \ginv{\psi}{N}' + iu\mathcal{H}\ginv{y}{N}) - 3\mathcal{H}'(\ginv{\psi}{N} + iu\ginv{y}{N}) - (\mathcal{H} + iu)\triangle\ginv{y}{N}] = 0\,.
\end{equation}
This equation takes the form of a screened Poisson equation for \(\ginv{y}{N}\). Isolating \(\ginv{y}{N}\), we have
\begin{equation}\label{eq:ysolvecn}
(\mathcal{H} + iu)\triangle\ginv{y}{N} + 3iu[\mathcal{H}' - \mathcal{H}(\mathcal{H} + iu)]\ginv{y}{N} = 3(\mathcal{H} + iu)(\mathcal{H}\ginv{\phi}{N} + \mathcal{H}\ginv{\psi}{N} - iu\ginv{\psi}{N} + \ginv{\psi}{N}') - 3\mathcal{H}'\ginv{\psi}{N}\,.
\end{equation}
We can then continue with the time part~\eqref{eq:timescalar} of the equations. This reads
\begin{equation}
\frac{1}{2}\kappa^2a^2\ginv{\mathcal{E}}{N} = f_T(\triangle\ginv{\psi}{N} - 3\mathcal{H}^2\ginv{\phi}{N} - 3\mathcal{H}\ginv{\psi}{N}' + 3u^2\ginv{\psi}{N}) + 12\frac{f_{TT}}{a^2}\mathcal{H}(\mathcal{H} + iu)^2(\triangle\ginv{y}{N} - 3\mathcal{H}\ginv{\phi}{N} - 3\ginv{\psi}{N}' + 3iu\ginv{\psi}{N})\,.
\end{equation}
One can substitute the spatial derivative of \(\ginv{y}{N}\) using the relation~\eqref{eq:ysolvecn} to obtain
\begin{equation}\label{eq:densvecn}
\frac{1}{2}\kappa^2a^2\ginv{\mathcal{E}}{N} = f_T(\triangle\ginv{\psi}{N} - 3\mathcal{H}^2\ginv{\phi}{N} - 3\mathcal{H}\ginv{\psi}{N}' + 3u^2\ginv{\psi}{N}) + 36\frac{f_{TT}}{a^2}\mathcal{H}(\mathcal{H} + iu)[\mathcal{H}(\mathcal{H} + iu) - \mathcal{H}'](\ginv{\psi}{N} + iu\ginv{y}{N})\,.
\end{equation}
One can then continue with the remaining mixed part~\eqref{eq:mixedscalar} of the equations, which reads
\begin{equation}
-\frac{1}{2}\kappa^2a^2(\bar{\rho} + \bar{p})\ginv{\mathcal{L}}{N} = f_T(\mathcal{H}\ginv{\phi}{N} + \ginv{\psi}{N}') + 12(\mathcal{H} + iu)[\mathcal{H}' - \mathcal{H}(\mathcal{H} + iu)]\frac{f_{TT}}{a^2}(\ginv{\psi}{N} + iu\ginv{y}{N})\,.
\end{equation}
Together with the previously found relation~\eqref{eq:densvecn} this yields
\begin{equation}
\frac{1}{2}\kappa^2a^2\ginv{\mathcal{E}}{C} = \frac{1}{2}\kappa^2a^2[\ginv{\mathcal{E}}{N} - 3\mathcal{H}(\bar{\rho} + \bar{p})\ginv{\mathcal{L}}{N}] = f_T(\triangle\ginv{\psi}{N} + 3u^2\ginv{\psi}{N})\,,
\end{equation}
where the left hand side is simply the density perturbation in the comoving gauge. Further, the off-diagonal symmetric equation
\begin{equation}
\kappa^2a^2\ginv{\mathcal{S}}{N} = f_T(\ginv{\psi}{N} - \ginv{\phi}{N}) - 12\frac{f_{TT}}{a^2}(\mathcal{H} + iu)[\mathcal{H}(\mathcal{H} + iu) - \mathcal{H}']\ginv{y}{N}
\end{equation}
yields the gravitational slip \(\ginv{\psi}{N} - \ginv{\phi}{N}\). Together with the trace of the spatial equations it yields
\begin{multline}
\frac{1}{2}\kappa^2a^2\left(\ginv{\mathcal{P}}{N} + \frac{2}{3}\triangle\ginv{\mathcal{S}}{N}\right) = f_T\left[\ginv{\psi}{N}'' + 2\mathcal{H}\ginv{\psi}{N}' + \mathcal{H}\ginv{\phi}{N}' + (\mathcal{H}^2 + 2\mathcal{H}')\ginv{\phi}{N} - u^2\ginv{\psi}{N}\right]\\
+ 4\frac{f_{TT}}{a^2}(\mathcal{H} + iu)\bigg\{3(\mathcal{H} + iu)\ginv{\psi}{N}'' + 3[3\mathcal{H}' - \mathcal{H}(\mathcal{H} + iu)]\ginv{\psi}{N}' + 3\mathcal{H}(\mathcal{H} + iu)\ginv{\phi}{N}' + 3iu(\mathcal{H}^2 - 3\mathcal{H}' + u^2)\ginv{\psi}{N}\\
+ 3[(5\mathcal{H} + 2iu)\mathcal{H}' - 2\mathcal{H}^2(\mathcal{H} + iu)]\ginv{\phi}{N} - (\mathcal{H} + iu)\triangle\ginv{y}{N}' - [2\mathcal{H}' + iu(\mathcal{H} + iu)]\triangle\ginv{y}{N}\bigg\}\\
+ 48\frac{f_{TTT}}{a^4}(\mathcal{H} + iu)^3[\mathcal{H}(\mathcal{H} + iu) - \mathcal{H}'](\triangle\ginv{y}{N} - 3\ginv{\psi}{N}' - 3\mathcal{H}\ginv{\phi}{N} + 3iu\ginv{\psi}{N})\,.
\end{multline}
After substituting \(\triangle\ginv{y}{N}\), this further simplifies to
\begin{multline}
\frac{1}{2}\kappa^2a^2\left(\ginv{\mathcal{P}}{N} + \frac{2}{3}\triangle\ginv{\mathcal{S}}{N}\right) = f_T\left[\ginv{\psi}{N}'' + 2\mathcal{H}\ginv{\psi}{N}' + \mathcal{H}\ginv{\phi}{N}' + (\mathcal{H}^2 + 2\mathcal{H}')\ginv{\phi}{N} - u^2\ginv{\psi}{N}\right]\\
+ 12\frac{f_{TT}}{a^2}\left[\mathcal{H}' - \mathcal{H}(\mathcal{H} + iu)\right](\mathcal{H} + iu)\left[2\ginv{\psi}{N}' + iu\ginv{y}{N}' - iu\ginv{\psi}{N} + (2\mathcal{H} + iu)\ginv{\phi}{N}\right]\\
+ \left\{12\frac{f_{TT}}{a^2}[(\mathcal{H}'' - 3\mathcal{H}'\mathcal{H})(\mathcal{H} + iu) + \mathcal{H}'^2 - iu\mathcal{H}(\mathcal{H} + iu)^2] + 144\frac{f_{TTT}}{a^4}[\mathcal{H}'(\mathcal{H} + iu) - \mathcal{H}(\mathcal{H} + iu)^2]^2\right\}(\ginv{\psi}{N} + iu\ginv{y}{N})\,.
\end{multline}
In this analysis, we observe that unlike the flat case, all the scalar modes are determined by the perturbed field equations at linear order, thus limiting the possibility of strongly coupled behavior for this background.

\subsubsection{Axial branch}\label{sssec:scalaxi}
In the axial case, see~\eqref{eq:axtet} and~\eqref{eq:axialbranch}, we find that the pseudo-scalar equation~\eqref{eq:pseudoscalar} takes the form
\begin{equation}\label{eq:pseudoscalaraxi}
-4uf_{TT}[u\triangle\ginv{\xi}{N} - \mathcal{H}\triangle\ginv{y}{N} + 3\mathcal{H}(\mathcal{H}^2 - \mathcal{H}' - u^2)\ginv{y}{N} + 3u^2\ginv{\psi}{N} + 3\mathcal{H}(\mathcal{H}\ginv{\phi}{N} + \ginv{\psi}{N}')] = 0\,,
\end{equation}
in the Newtonian gauge, so that the pseudo-scalar perturbation \(\ginv{\xi}{N}\) is coupled to the scalar perturbations. Together with the antisymmetric equation~\eqref{eq:asymscalar}, which reads
\begin{equation}\label{eq:ypoissaxin}
f_{TT}\mathcal{H}[3\mathcal{H}(\mathcal{H}\ginv{\phi}{N} + \mathcal{H}\ginv{\psi}{N} + \ginv{\psi}{N}' - u\mathcal{H}\ginv{\xi}{N}) - 3\mathcal{H}'(\ginv{\psi}{N} - u\ginv{\xi}{N}) - \mathcal{H}\triangle\ginv{y}{N} + 3u^3\ginv{\xi}{N} + u\triangle\ginv{\xi}{N}] = 0\,,
\end{equation}
we can eliminate the spatial derivatives, and are left with the purely algebraic equation
\begin{equation}
12uf_{TT}(\mathcal{H}^2 - \mathcal{H}' - u^2)(\ginv{\psi}{N} - u\ginv{\xi}{N} - \mathcal{H}\ginv{y}{N}) = 0\,.
\end{equation}
This can now be solved for the pseudo-scalar \(\ginv{\xi}{N}\). Substituting back into the original equations yields a screened Poisson equation for \(\ginv{y}{N}\), which reads
\begin{equation}
\mathcal{H}[2\triangle\ginv{y}{N} + 3(\mathcal{H}' - \mathcal{H}^2 + u^2)\ginv{y}{N}] = \triangle\ginv{\psi}{N} + 3u^2\ginv{\psi}{N} + 3\mathcal{H}\ginv{\psi}{N}' + 3\mathcal{H}^2\ginv{\phi}{N}\,.
\end{equation}
We then continue with the time component~\eqref{eq:timescalar}, from which we obtain
\begin{equation}
\frac{1}{2}\kappa^2a^2\ginv{\mathcal{E}}{N} = f_T(\triangle\ginv{\psi}{N} - 3\mathcal{H}^2\ginv{\phi}{N} - 3\mathcal{H}\ginv{\psi}{N}' + 3u^2\ginv{\psi}{N}) + 12\frac{f_{TT}}{a^2}\mathcal{H}^2(\mathcal{H}\triangle\ginv{y}{N} - u\triangle\ginv{\xi}{N} - 3\mathcal{H}^2\ginv{\phi}{N} - 3\mathcal{H}\ginv{\psi}{N}' - 3u^2\ginv{\psi}{N})\,.
\end{equation}
After substituting \(\ginv{\xi}{N}\) and \(\triangle\ginv{y}{N}\), this yields
\begin{equation}\label{eq:densaxin}
\frac{1}{2}\kappa^2a^2\ginv{\mathcal{E}}{N} = f_T(\triangle\ginv{\psi}{N} - 3\mathcal{H}^2\ginv{\phi}{N} - 3\mathcal{H}\ginv{\psi}{N}' + 3u^2\ginv{\psi}{N}) + 36\frac{f_{TT}}{a^2}\mathcal{H}^3(\mathcal{H}^2 - \mathcal{H}' - u^2)\ginv{y}{N}\,.
\end{equation}
By combining this equation with the remaining mixed part~\eqref{eq:mixedscalar}, which reads
\begin{equation}
-\frac{1}{2}\kappa^2a^2(\bar{\rho} + \bar{p})\ginv{\mathcal{L}}{N} = f_T(\mathcal{H}\ginv{\phi}{N} + \ginv{\psi}{N}') + 12\mathcal{H}(\mathcal{H}' - \mathcal{H}^2 + u^2)\frac{f_{TT}}{a^2}(\ginv{\psi}{N} - u\ginv{\xi}{N})\,,
\end{equation}
we find the expression
\begin{equation}
\frac{1}{2}\kappa^2a^2\ginv{\mathcal{E}}{C} = \frac{1}{2}\kappa^2a^2[\ginv{\mathcal{E}}{N} - 3\mathcal{H}(\bar{\rho} + \bar{p})\ginv{\mathcal{L}}{N}] = f_T(\triangle\ginv{\psi}{N} + 3u^2\ginv{\psi}{N})\,,
\end{equation}
where we have expressed the left hand side in the comoving gauge. Note that this equation is identical to the vector branch. Continuing with the symmetric off-diagonal equation, we find the result
\begin{equation}
\kappa^2a^2\ginv{\mathcal{S}}{N} = f_T(\ginv{\psi}{N} - \ginv{\phi}{N}) - 12\frac{f_{TT}}{a^2}\mathcal{H}(\mathcal{H}^2 - \mathcal{H}' - u^2)\ginv{y}{N}\,,
\end{equation}
which determines the gravitational slip \(\ginv{\psi}{N} - \ginv{\phi}{N}\). Together with the trace of the spatial equations it yields
\begin{multline}
\frac{1}{2}\kappa^2a^2\left(\ginv{\mathcal{P}}{N} + \frac{2}{3}\triangle\ginv{\mathcal{S}}{N}\right) = f_T\left[\ginv{\psi}{N}'' + 2\mathcal{H}\ginv{\psi}{N}' + \mathcal{H}\ginv{\phi}{N}' + (\mathcal{H}^2 + 2\mathcal{H}')\ginv{\phi}{N} - u^2\ginv{\psi}{N}\right]\\
+ 4\frac{f_{TT}}{a^2}\bigg[3\mathcal{H}^2\ginv{\psi}{N}'' + 3\mathcal{H}(3\mathcal{H}' - \mathcal{H}^2 + u^2]\ginv{\psi}{N}' + 3\mathcal{H}^3\ginv{\phi}{N}' + 3u^2(\mathcal{H}' - u^2)\ginv{\psi}{N}\\
+ 3\mathcal{H}^2(5\mathcal{H}' - 2\mathcal{H}^2 + u^2)\ginv{\phi}{N} - \mathcal{H}^2\triangle\ginv{y}{N}' - \mathcal{H}(2\mathcal{H}' - u^2)\triangle\ginv{y}{N} + u\mathcal{H}\triangle\ginv{\xi}{N}' + u(\mathcal{H}' - u^2)\triangle\ginv{\xi}{N}\bigg]\\
+ 48\frac{f_{TTT}}{a^4}\mathcal{H}^2(\mathcal{H}^2 - \mathcal{H}' - u^2)(\mathcal{H}\triangle\ginv{y}{N} - 3\mathcal{H}\ginv{\psi}{N}' - 3\mathcal{H}^2\ginv{\phi}{N} - 3u^2\ginv{\psi}{N} - u\triangle\ginv{\xi}{N})\,.
\end{multline}
This can further be simplified by substituting \(\ginv{\xi}{N}\) and \(\triangle\ginv{y}{N}\) from the previous equations, so that one obtains
\begin{multline}
\frac{1}{2}\kappa^2a^2\left(\ginv{\mathcal{P}}{N} + \frac{2}{3}\triangle\ginv{\mathcal{S}}{N}\right) = f_T\left[\ginv{\psi}{N}'' + 2\mathcal{H}\ginv{\psi}{N}' + \mathcal{H}\ginv{\phi}{N}' + (\mathcal{H}^2 + 2\mathcal{H}')\ginv{\phi}{N} - u^2\ginv{\psi}{N}\right] + 12\frac{f_{TT}}{a^2}\mathcal{H}(\mathcal{H}' - \mathcal{H}^2 + u^2)(\ginv{\psi}{N}' + 2\mathcal{H}\ginv{\phi}{N} + \mathcal{H}\ginv{y}{N}')\\
+ \left\{12\frac{f_{TT}}{a^2}\mathcal{H}[\mathcal{H}\mathcal{H}'' + (2\mathcal{H}' - 4\mathcal{H}^2 + u^2)\mathcal{H}' + u^2\mathcal{H}^2 - u^4] + 144\frac{f_{TTT}}{a^4}\mathcal{H}^3(\mathcal{H}' - \mathcal{H}^2 + u^2)^2\right\}\ginv{y}{N}\,.
\end{multline}
This prescribes the behaviour of all modes in this branch of the perturbative sector.

\subsubsection{Flat case}\label{sssec:scalflat}
We finally come to the flat limiting case. In this case we find that the pseudo-scalar equation~\eqref{eq:pseudoscalar} is solved identically. This can also be seen from the corresponding equations~\eqref{eq:pseudoscalarvec} and~\eqref{eq:pseudoscalaraxi} in the vector and axial branches, which vanish for \(u \to 0\). Hence, in the flat case, the pseudo-scalar perturbation \(\ginv{\xi}{N}\) is not determined from the field equations, and yields a remnant symmetry of the linearized field equations around the flat FLRW background. The remaining equations possess as similar structure as for the spatially curved background. From the mixed antisymmetric equations~\eqref{eq:asymscalar} one obtains
\begin{equation}
f_{TT}\mathcal{H}[3\mathcal{H}(\mathcal{H}\ginv{\phi}{N} + \mathcal{H}\ginv{\psi}{N} + \ginv{\psi}{N}') - 3\mathcal{H}'\ginv{\psi}{N} - \mathcal{H}\triangle\ginv{y}{N}] = 0\,,
\end{equation}
which is a Poisson equation for \(\ginv{y}{N}\), and can be used to eliminate \(\triangle\ginv{y}{N}\) in the remaining equations. For the time component~\eqref{eq:timescalar}, this yields
\begin{equation}
\begin{split}
\frac{1}{2}\kappa^2a^2\ginv{\mathcal{E}}{N} &= f_T\triangle\ginv{\psi}{N} - 3\left(f_T + 12\mathcal{H}^2\frac{f_{TT}}{a^2}\right)\mathcal{H}(\mathcal{H}\ginv{\phi}{N} + \ginv{\psi}{N}') + 12\frac{f_{TT}}{a^2}\mathcal{H}^3\triangle\ginv{y}{N}\\
&= f_T(\triangle\ginv{\psi}{N} - 3\mathcal{H}\ginv{\psi}{N}' - 3\mathcal{H}^2\ginv{\phi}{N}) + 36\frac{f_{TT}}{a^2}\mathcal{H}^2(\mathcal{H}^2 - \mathcal{H}')\ginv{\psi}{N}\,.
\end{split}
\end{equation}
Together with the remaining mixed part~\eqref{eq:mixedscalar}, which reads
\begin{equation}
-\frac{1}{2}\kappa^2a^2(\bar{\rho} + \bar{p})\ginv{\mathcal{L}}{N} = f_T(\mathcal{H}\ginv{\phi}{N} + \ginv{\psi}{N}') + 12\mathcal{H}(\mathcal{H}' - \mathcal{H}^2)\frac{f_{TT}}{a^2}\ginv{\psi}{N}\,,
\end{equation}
this combines to
\begin{equation}
\frac{1}{2}\kappa^2a^2\ginv{\mathcal{E}}{C} = \frac{1}{2}\kappa^2a^2[\ginv{\mathcal{E}}{N} - 3\mathcal{H}(\bar{\rho} + \bar{p})\ginv{\mathcal{L}}{N}] = f_T\triangle\ginv{\psi}{N}\,,
\end{equation}
which is simply the limit \(u \to 0\) of what we have found for the spatially curved background. Finally, we have the off-diagonal equation
\begin{equation}
\kappa^2a^2\ginv{\mathcal{S}}{N} = f_T(\ginv{\psi}{N} - \ginv{\phi}{N}) - 12\frac{f_{TT}}{a^2}\mathcal{H}(\mathcal{H}^2 - \mathcal{H}')\ginv{y}{N}\,,
\end{equation}
which combines with the trace into
\begin{multline}
\frac{1}{2}\kappa^2a^2\left(\ginv{\mathcal{P}}{N} + \frac{2}{3}\triangle\ginv{\mathcal{S}}{N}\right) = f_T\left[\ginv{\psi}{N}'' + 2\mathcal{H}\ginv{\psi}{N}' + \mathcal{H}\ginv{\phi}{N}' + (\mathcal{H}^2 + 2\mathcal{H}')\ginv{\phi}{N}\right] + 48\frac{f_{TTT}}{a^4}\mathcal{H}^3(\mathcal{H}^2 - \mathcal{H}')(\triangle\ginv{y}{N} - 3\ginv{\psi}{N}' - 3\mathcal{H}\ginv{\phi}{N})\\
+ 4\frac{f_{TT}}{a^2}\mathcal{H}\left[3\mathcal{H}\ginv{\psi}{N}'' + 3(3\mathcal{H}' - \mathcal{H}^2)\ginv{\psi}{N}' + 3\mathcal{H}^2\ginv{\phi}{N}' + 3\mathcal{H}(5\mathcal{H}' - 2\mathcal{H}^2)\ginv{\phi}{N} - \mathcal{H}\triangle\ginv{y}{N}' - 2\mathcal{H}'\triangle\ginv{y}{N}\right]\,.
\end{multline}
Also here we can eliminate \(\triangle\ginv{y}{N}\) to obtain
\begin{multline}
\frac{1}{2}\kappa^2a^2\left(\ginv{\mathcal{P}}{N} + \frac{2}{3}\triangle\ginv{\mathcal{S}}{N}\right) = f_T\left[\ginv{\psi}{N}'' + 2\mathcal{H}\ginv{\psi}{N}' + \mathcal{H}\ginv{\phi}{N}' + (\mathcal{H}^2 + 2\mathcal{H}')\ginv{\phi}{N}\right] + 144\frac{f_{TTT}}{a^4}\mathcal{H}^2(\mathcal{H}^2 - \mathcal{H}')^2\ginv{\psi}{N}\\
+ 12\frac{f_{TT}}{a^2}\left[2\mathcal{H}(\mathcal{H}' - \mathcal{H}^2)(\ginv{\psi}{N}' + \mathcal{H}\ginv{\phi}{N}) + (\mathcal{H}\mathcal{H}'' + \mathcal{H}'^2 - 3\mathcal{H}^2\mathcal{H}')\ginv{\psi}{N}\right]\,.
\end{multline}

Thus, for the scalar perturbations we found that not all of them are determined in all three cases discussed. In the flat case the pseudo-scalar perturbation is undetermined. Hence, if its appears in higher order perturbation theory its arbitrary value makes it impossible to solve these, which poses a problem to the predictability of the theory. This feature is the strong coupling problem since the Hamiltonian formalism suggests that there should be one additional degree of freedom than GR.

\section{Perturbative degrees of freedom}\label{sec:dof}
We now study the obtained perturbative field equations under the aspect of counting the dynamical degrees of freedom which are present in the perturbations. Already in section~\ref{ssec:tensorpert} we have seen two propagating tensor modes around each of the cosmological background branches, while in section~\ref{ssec:vec_pert} we have seen that no further vector modes appear in the spatially curved FLRW background, as compared to the flat FLRW case. We therefore devote this discussion entirely to the scalar perturbations, whose governing equations we derived in section~\ref{ssec:scal_pert}, and where we see a qualitative change in the rank of the linear system of equations.

If we collectively denote the (pseudo-)scalar perturbations by \(\vek{X} = (\ginv{\phi}{N}, \ginv{\psi}{N}, \ginv{y}{N}, \ginv{\xi}{N})\) and the right hand side of the scalar equations~\eqref{eq:pertscalsep}, which constitutes the matter source, by \(\vek{Y}\), and perform a harmonic expansion as discussed in section~\ref{ssec:harmonic} to replace the Laplace operator \(\triangle\) by the corresponding eigenvalue \(-k^2\), the equations take the schematic form
\begin{equation}
\mat{M}_2\vek{X}'' + \mat{M}_1\vek{X}' + \mat{M}_0\vek{X} = \vek{Y}\,,
\end{equation}
where the $(6 \times 4)$-matrices \(\mat{M}_{0,1,2}\) depend on the dynamical background geometry and the eigenvalue \(k^2\). Note that this system is consistent despite the fact that it contains six equations for four variables, since the right hand side is subject to the constraints arising from energy-momentum conservation. To further analyse this system, we write it in the form
\begin{equation}\label{eq:firstorder}
\mat{\tilde{M}}_1\vek{\tilde{X}}' + \mat{\tilde{M}}_0\vek{\tilde{X}} = \begin{pmatrix}
\mat{0} & \mat{1}\\
\mat{M}_2 & \mat{0}
\end{pmatrix} \cdot \begin{pmatrix}
\vek{X}''\\
\vek{X}'
\end{pmatrix} + \begin{pmatrix}
-\mat{1} & \mat{0}\\
\mat{M}_1 & \mat{M}_0
\end{pmatrix} \cdot \begin{pmatrix}
\vek{X}'\\
\vek{X}
\end{pmatrix} = \begin{pmatrix}
\vek{0}\\
\vek{Y}
\end{pmatrix} = \vek{\tilde{Y}}
\end{equation}
as a first order system. By successively performing Gaussian elimination, the combined block matrix \(\begin{pmatrix} \mat{\tilde{M}}_1 & \mat{\tilde{M}}_0 \end{pmatrix}\) can be brought into row echelon form, so that the equivalent system schematically reads
\begin{equation}
\begin{pmatrix}
\mat{D}_1 & \mat{D}_0\\
\mat{0} & \mat{A}\\
\mat{0} & \mat{0}
\end{pmatrix} \cdot \begin{pmatrix}
\vek{\tilde{X}}'\\
\vek{\tilde{X}}
\end{pmatrix} = \begin{pmatrix}
\vek{Y}_D\\
\vek{Y}_A\\
\vek{0}
\end{pmatrix}\,.
\end{equation}
Denoting the number of rows in these blocks by \(N_D, N_A, N_I\), we find that the lowermost block contains \(N_I\) equations which are satisfied identically, followed by \(N_A\) purely algebraic equations or constraints and finally \(N_D\) differential equations. It follows from the construction of these blocks that the $(N_A \times 8)$-matrix \(\mat{A}\) is of maximal rank, which is given by \(N_A\). The space of solutions is thus of dimension \(8 - N_A\), and can be written as
\begin{equation}
\mat{A} \cdot (\mat{S} \cdot \vek{V} + \vek{W}) = \vek{Y}_A\,,
\end{equation}
where \(\vek{W}\) denotes a particular solution satisfying \(\mat{A} \cdot \vek{W} = \vek{Y}_A\), the columns of \(\mat{S}\) span the kernel of \(\mat{A}\) and \(\vek{V}\) is a vector of \(8 - N_A\) arbitrary coefficients. Inserting
\begin{equation}
\tilde{X} = \mat{S} \cdot \vek{V} + \vek{W}
\end{equation}
into the block of differential equations then gives another set of \(N_D\) first order differential equations in the remaining variables \(\vek{V}\). By repeatedly performing the same steps as for the initial first order system~\eqref{eq:firstorder}, one can find and solve all constraints, until one is left with a system of differential equations only and no algebraic equations remain. The number of equations in this final system determines the number of initial conditions which must be supplied in order to solve the equations.

We then apply this algorithm to the scalar perturbation equations derived in section~\ref{ssec:scal_pert}, where we consider the generic case, i.e., we assume that all appearing matrices have the maximal possible rank, which is not further reduced by a particular form of the function \(f\) or the cosmological background evolution. We then find the following results:
\begin{enumerate}
\item
For the two spatially curved cases discussed in sections~\ref{sssec:scalvec} and~\ref{sssec:scalaxi} we find at the first step \(N_I = 0\) trivial equations, \(N_A = 5\) algebraic equations and \(N_D = 5\) differential equations. Solving the algebraic equations, we are thus left with a solution space of dimension \(3\). Inserting this solution into the remaining differential equations and performing Gaussian elimination again, we find \(N_I = 2\) identically satisfied equations, \(N_A = 3\) algebraic equations and \(N_D = 0\) remaining differential equations. The system is fully constrained.

\item
In the spatially flat case discussed in section~\ref{sssec:scalflat}, we see that the vanishing curvature parameter \(u = 0\) leads to a reduced rank of the matrices constituting the linear differential system. At the first step, we find \(N_I = 1\) identically satisfied equation, which arises from the fact that the pseudo-scalar equation is identically satisfied and the perturbation \(\ginv{\xi}{N}\) is undetermined. There are \(N_A = 4\) further algebraic equations and \(N_D = 5\) differential equations left. Solving the former gives a solution space of dimension \(4\). Inserting this into the remaining differential system, we have \(N_I = 2\) identically satisfied equations, \(N_A = 2\) algebraic equations and \(N_D = 1\) differential equations. The algebraic equations constrain the system further to only two remaining variables, \(\ginv{\xi}{N}\) and \(\ginv{\xi}{N}'\), which are set in relation by the final differential equation. Hence, we find that \(\ginv{\xi}{N}\) is undetermined (and fixes its time derivative \(\ginv{\xi}{N}'\)), while all other scalar perturbations are fully constrained.
\end{enumerate}
We finally remark that in the TEGR case \(f(T) = T\) both \(\ginv{\xi}{N}\) and \(\ginv{y}{N}\) are undetermined, and one is left with the two Bardeen potentials \(\ginv{\phi}{N}\) and \(\ginv{\psi}{N}\), as one would expect.

\section{Conclusions}\label{sec:conclu}

The main pillar of modern cosmology is a homogeneous and isotropic FLRW Universe, which serves as background stage for the propagation of perturbations which are the source for many properties of the cosmos we observe. Without the underlying FLRW geometry it would be exceedingly difficult to make any predictions from theoretical models.

In this work, we extend the study of the evolution of perturbations in $f(T)$ gravity, in the covariant formulation outlined in Section \ref{sec:f_T_Intro}, to non-flat cosmologies on the one hand to be able to address recent cosmological observations which may have a slight preference for a closed Universe \cite{DiValentino:2019qzk}, while on the other hand, to demonstrate explicitly the existence or absence of the strong coupling issue, see \cite{Golovnev:2020zpv}, around the one or the other geometric background solution of the theory.

Our main result is that for spatially curved homogeneous and isotropic teleparallel background geometries in $f(T)$-gravity the strong coupling problem continues to appear due to the non-propagation nature of some scalar perturbation modes, in addition to being present for the flat case. This extends the result from the flat case to any non-flat FLRW cosmology. However, it may be the case that certain other cosmologies can evade this result and not express the strong coupling issue.

There has been a long discussion regarding the number of degrees of freedom in $f(T)$ gravity. The Hamiltonian formalism has been studied by several authors (see a review~\cite{Blixt:2020ekl}) and found that the number of degrees of freedom is either three or five~\cite{Ferraro:2018tpu,Blagojevic:2020dyq}. This number depends on the tetrad and the symmetries imposed for the torsion scalar $T$ (there are two branches in the Hamiltonian formalism). When the torsion scalar depends only on time (as in FLRW cosmologies), the expected number of degrees of freedom predicted by the Hamiltonian formalism is three. On the other hand, by taking perturbations around flat FLRW, no new modes, i.e. just two as GR, appear in $f(T)$ gravity~\cite{Golovnev:2018wbh}. This suggests that this theory is strongly coupled against flat FLRW cosmology.

Our strategy to reach this conclusion was to consider the two most general non-flat homogeneous and isotropic tetrads, the axial and the vector branch displayed in Eqs.~\eqref{eq:vectet} and \eqref{eq:axtet}, and to explore their background and perturbative evolution in $f(T)$ gravity to study whether it contains any strongly coupled modes, or not. Both branches converge smoothly to the same flat FLRW  limit as the curvature parameter tends to zero. The background evolution of the two branches in $f(T)$ gravity is governed by the axial and vector Friedmann equation shown in Eqs.~\eqref{eq:back_FE_vector} and~\eqref{eq:back_FE_axial}. Our main interest lies in the evolution of the perturbations on these homogeneous and isotropic backgrounds. To study linear perturbation theory in teleparallel gravity we used a $3+1$ and a differential decomposition of all ingredients: for the degrees of freedom in the linear perturbation theory, which are determined from the tetrad perturbations, this decomposition is displayed in Eq.~\eqref{eq:cosmoperttet}, for the perturbations of the energy-momentum tensor it can be found in Eq.~\eqref{eq:cosmopertem2} and for the field equations it leads to six scalar \eqref{eq:pertscalgen}, four vector \eqref{eq:pertvecgen} and a tensorial \eqref{eq:perttengen} linearized field equations.

In Sec.~\ref{sec:cosmo_perturb} we assembled all these ingredients together to investigate the evolution of the perturbations from the perturbed field equations in the context of the non-flat background cosmology. Starting from the background Friedmann equations for the vector \eqref{eq:back_FE_vector} and axial \eqref{eq:back_FE_axial} branch we study different perturbation sectors (vector/axial/flat branch - scalar/vector/tensor perturbations) in turn to explore the fate of the perturbative degrees of freedom in each case.
\begin{itemize}
    \item The tensor perturbations are given for the vector and the axial branch in Eqs.~\eqref{eq:gwpe_vec} and \eqref{eq:gwpe_ax} respectively. They both predict the propagating of these modes with the speed of light and contain an additional term proportional to $f_{TT}$ compared to the flat case. Most importantly, all tensor modes are well determined by the perturbative equations displayed in Section \ref{ssec:tensorpert}.
    \item A similar conclusion can be drawn for the vector perturbations. They are fully determined by equations \eqref{eq:vecvec1} to \eqref{eq:vecvec3} and \eqref{eq:vecax1} to \eqref{eq:vecax4}. Actually both cases of cosmological curvature result in a vector sector that does not evolve and so does not contribute to the cosmology of the theory, as discussed in Section \ref{ssec:vec_pert}.
    \item Finally, for the pseudo-scalar and scalar sector the situation is different. This is the sector that exhibits strong coupling in the flat case. Equations \eqref{eq:pseudoscalarvec} and~\eqref{eq:pseudoscalaraxi}, which determine the value of the pseudo-scalar mode $\ginv{\xi}{N}$ in the non-flat case, are identically zero in the flat case. For the flat case, $\ginv{\xi}{N}$ is undetermined. Thus, if this mode couples to further modes in higher order perturbations theory, or in the full theory, the theory is not predictive, since this mode can assume any value.

    For all other scalar modes this problem does not emerge, as we explained in Section \ref{ssec:scal_pert}.
\end{itemize}

Hence, we find that in spatially curved $f(T)$ cosmology, all perturbation modes are determined at the linear level, but some of the modes are non-propagating and so the strong coupling issue remains present in this setting.

An interesting future research direction is to perform the analogue analysis for $f(Q)$ non-metricity theories of gravity~\cite{BeltranJimenez:2019tme,Hohmann:2021ast} to identify the first insights about the existence or non existence of a strong coupling problem. For a definite answer such an analysis must then be complemented by a Hamiltonian analysis of $f(Q)$-gravity, which is nowadays unexplored~\cite{DAmbrosio:2020nqu}.

\begin{acknowledgments}
The authors would like to acknowledge important discussions on the manuscript with Daniel Blixt, Alexey Golovnev, María José Guzmán, Jose Beltrán Jimenéz and Tomi Sebastian Koivisto. C.P. was funded by the Deutsche Forschungsgemeinschaft (DFG, German Research Foundation) - Project Number 420243324 and acknowledges support from the DFG funded cluster of excellence Quantum Frontiers. M.H. acknowledges support by the Estonian Ministry for Education and Science through the Personal Research Funding Grants PRG356, as well as the European Regional Development Fund through the Center of Excellence TK133 ``The Dark Side of the Universe''. S.B. is supported by JSPS Postdoctoral Fellowships for Research in Japan and KAKENHI Grant-in-Aid for Scientific Research No. JP21F21789. The authors would like to acknowledge networking support by the COST Action CA18108. The work was supported by Nazarbayev University Faculty Development Competitive Research Grant No. 11022021FD2926. This research work was supported by the Hellenic Foundation for Research and Innovation (H.F.R.I.) under the “First Call for H.F.R.I. Research Projects to support Faculty members and Researchers and the procurement of high-cost research equipment grant” (Project Number: 2251).
\end{acknowledgments}

\bibliographystyle{utphys}
\bibliography{references}

\end{document}